\documentclass[usegraphicx,useAMS,usenatbib]{mn2e}

\newcommand{\beq}{\begin{equation}}
\newcommand{\eeq}{\end{equation}}
\newcommand{\beqa}{\begin{eqnarray}}
\newcommand{\eeqa}{\end{eqnarray}}

\def\lexp{\mathop{{\bigl\langle}}\nolimits}
\def\rexp{\mathop{{\bigl\rangle}}\nolimits}
\def\Lexp{\mathop{{\Bigl\langle}}\nolimits}
\def\Rexp{\mathop{{\Bigl\rangle}}\nolimits}
\def\Dlexp{\mathop{{\bigl\langle\!\bigl\langle}}\nolimits}
\def\Drexp{\mathop{{\bigl\rangle\!\bigr\rangle}}\nolimits}

\def\d{\rmn{d}}
\def\e{\rmn{e}}

\def\bm{\bmath}

\def\r{{\bm{r}}}

\def\Mbar{{\bar M}}
\def\Nbar{{\bar N}}
\def\nbar{{\bar n}}
\def\rhobar{{\bar\rho}}
\def\xibar{{\bar\xi}}
\def\mubar{{\bar\mu}}
\def\bbar{{\bar b}}

\def\M{m}

\def\tsum{\mathop{{\textstyle \sum}}}

\def\min{\rmn{min}}
\def\max{\rmn{max}}
\def\Msun{{M_{\sun}}}
\def\Mpc{\,{\rmn{Mpc}}}
\def\fit{\rmn{f}}
\def\model{\rmn{m}}
\def\grid{\rmn{grid}}
\def\tot{\rmn{tot}}
\def\NFW{\rmn{NFW}}
\def\ISO{\rmn{ISO}}

\def\RAMSES{{\tt RAMSES}}
\def\adaptaHOP{{\tt adaptaHOP}}

\begin{document}

\title[Cell Count Moments in the Halo Model]
{Cell Count Moments in the Halo Model}

\author[J. N. Fry et al.]
{J. N. Fry,$^{1,2}$\thanks{E-mail: 
fry@phys.ufl.edu (JNF); colombi@iap.fr (SC); fosalba@ieec.uab.es (PF); 
abalaraman@georgiasouthern.edu (AB); szapudi@ifa.hawaii.edu (IS); 
romain.teyssier@cea.fr (RT)}
S. Colombi,$^{2\star} $ Pablo Fosalba,$^{3\star}$ 
Anand Balaraman,$^{1,4\star}$ Istv\'{a}n Szapudi$^{5\star}$ 
\newauthor
and R. Teyssier$^{6,7\star}_{\strut}$\\
$^1$Department of Physics, University of Florida, 
Gainesville FL 32611-8440, USA \\
$^2$Institut d'Astrophysique de Paris \& UPMC (UMR 7095), 
98 bis boulevard Arago, 75014 Paris, France \\
$^3$Institut de Ci\`{e}ncies de l'Espai, IEEC--CSIC, 
Campus UAB, F. de Ci\`{e}ncies, Torre C5 par-2, Barcelona 08193, Spain \\
$^4$Department of Physics, Georgia Southern University, 
Statesboro GA 30460, USA \\
$^5$Institute for Astronomy, University of Hawaii, 
2680 Woodlawn Drive, Honolulu, HI 96822, USA \\
$^6$
Service d'Astrophysique, CEA Saclay, Orme des Merisiers, 
91191 Gif-sur-Yvette Cedex, France \\
$^7$Institut f\"{u}r Theoretische Physik, Universit\"{a}t Z\"{u}rich, 
Winterthurer Strasse 190, CH-8057 Z\"{u}rich, Schweiz 
}

\maketitle

\begin{abstract}

We study cell count moments up to fifth order of the distributions 
of haloes, of halo substructures as a proxy for galaxies, and of mass 
in the context of the halo model and compare theoretical predictions 
to the results of numerical simulations.
On scales larger than the size of the largest cluster, we present  
a simple point cluster model in which results depend only on 
cluster-cluster correlations and on the distribution of the number 
of objects within a cluster, or cluster occupancy.
The point cluster model leads to expressions for moments of galaxy 
counts in which the volume-averaged moments on large scales approach 
those of the halo distribution and on smaller scales exhibit 
hierarchical clustering with amplitudes $S_k$ determined by moments 
of the occupancy distribution.
In this limit, the halo model predictions are purely combinatoric, 
and have no dependence on halo profile, concentration parameter, 
or potential asphericity.
The full halo model introduces only two additional effects: 
on large scales, haloes of different mass have different clustering 
strengths, introducing relative bias parameters; and on the smallest 
scales, halo structure is resolved and details of the halo profile 
become important, introducing shape-dependent form factors.
Because of differences between discrete and continuous statistics, 
the hierarchical amplitudes for galaxies and for mass behave differently 
on small scales even if galaxy number is exactly proportional to mass, 
a difference that is not necessarily well described in terms of bias.

\end{abstract}


\section{Introduction}

Describing the properties of the distribution of matter in the 
universe in terms of the masses, spatial distribution, and shapes 
of clusters, or haloes, is an enterprise with a long history 
\citep{NS52,MS77,P80,SB91,SS94,S96b}.
Recently, with the new ingredient of a universal halo profile 
found in numerical simulations \citep{NFW97,MQGSL99,NHPJFWSSQ04}, 
interest in the model has been rekindled \citep{S00,MF00b,PS00,SSHJ01}.
This model is not seen as literally true, but its constructions 
give plausible estimates for correlation functions because 
at a given scale, density-weighted statistics are dominated by 
the highest density systems, the collapsed haloes.
The model has been shown to reproduce two-point and higher order 
density correlation functions in simulations, and, with a carefully  
chosen halo mass function and `concentration parameter,' can be consistent 
with self-similar stable clustering \citep{MF00a,SPJWFTPEC03}.
Among its many other applications to weak gravitational lensing, 
pair velocities, the Ly-$\alpha$ forest, and CMB foregrounds, we find 
that the halo model also allows us to address the different behaviors 
of the continuous mass density and of discrete objects such as galaxies.

In this paper we reexamine the statistical behavior of integral moments 
of total mass or number counts in the context of the halo model.
Our results, formulated directly in position space, 
complement and extend those of \citet{SSHJ01} in the Fourier domain.
In Section 2 we present definitions of the various statistics we need 
and introduce generating function tools that will be applied in later sections.
In Section 3 we apply the probability generating function machinery 
for a system of identical clusters in the point cluster limit, 
a model we call the `naive halo model,' to express the statistics 
of counts in cells in terms of properties of the halo number 
and halo occupancy distributions.
In Section 4 we compare the model to results obtained from numerical 
simulations.
We find that the naive point cluster model describes the 
qualitative behavior but fails in quantitative detail, but insight 
gathered from the model in the generating function formalism 
is easily applied in the full halo model.
This leads us in Section 5 to consider the halo model in its full detail, 
summing over haloes of different mass, with both halo occupations 
and halo correlations functions of halo mass.
Properly interpreted, the naive point cluster results 
obtained using the generating function continue to apply 
when averaged over halo masses and over galaxy positions within a halo.
This allows us to extend to small scales, where haloes are resolved, 
introducing geometric form factors for haloes 
that can no longer be considered as points.
Working directly in space instead of in the Fourier transform domain 
allows us to exhibit manifestly symmetries under particle exchange 
at all orders; avoids the necessity to introduce an approximate 
factoring of window function products $ W_1 W_2 W_{12} \approx W_1^2 W_2^2 $, 
etc.; and avoids the necessity to make any assumptions or approximations 
about configuration dependence.
Known forms of the halo mass function, bias, and occupation number 
allow us to compute from first principles results 
in scale-free and specific cosmological models.
Section 6 contains a final discussion, 
and some technical results are included in appendices.

\section{Statistical Definitions}

We study statistics of the continuous mass density and of 
discrete objects that for convenience we denote as ``galaxies.''
For a galaxy number distribution that is a random sampling of 
a process with a smooth underlying number density field $ n(\r) $, 
factorial moments of the number of galaxies in a randomly placed volume 
directly reflect moments of the underlying continuous density field 
\citep{SS93}, 
\beq
\lexp N^{[k]} \rexp = \Nbar^k \, \mubar_k , 
\eeq
where $ N^{[k]} = N! / (N-k)! = N(N-1) \cdots (N-k+1) $, 
and the moments $ \mubar_k $ are volume averages of corresponding 
moments of the underlying density field, 
\beqa
&&\mubar_k = {1 \over V^k} \int_V \d^3r_1 \cdots \d^3r_k \, 
\mu_k(\r_1, \dots, \r_k) , \\
&&\lexp n(\r_1) \cdots n(\r_k) \rexp = \nbar^k \, \mu_k(\r_1,\dots,\r_k) , 
\eeqa
typically integrated over a spherical volume of radius $R$.
Moments of powers $ \lexp N^k \rexp $ then contain contributions 
arising from discreteness; for $ k = 2 $ through 5 these are 
\beqa
\lexp N^2 \rexp &=& \Nbar + \Nbar^2 \mubar_2 , \label{disc2} \\
\lexp N^3 \rexp &=& \Nbar + 3  \Nbar^2 \mubar_2 + \Nbar^3 \mubar_3 , \\
\lexp N^4 \rexp &=& \Nbar + 7 \Nbar^2 \mubar_2 
 + 6 \Nbar^3 \mubar_3 + \Nbar^4 \mubar_4 , \\
\lexp N^5 \rexp &=& \Nbar  + 15 \Nbar^2 \mubar_2 + 25 \Nbar^3 \mubar_3
  + 10 \Nbar^4 \mubar_4 + \Nbar^5 \mubar_5  . \label{disc5}
\eeqa
In the limit $ \Nbar = \lexp N \rexp \gg 1 $ the highest power of 
$\Nbar$ dominates and $ \lexp N^k \rexp = \Nbar^k \, \mubar_k $, 
as for a continuous density; 
the factorial moment, in removing the lower order or discreteness terms, 
leaves a discreteness corrected moment that reflects only spatial clustering.
The moments $ \mubar_k $ can be additionally separated into 
irreducible contributions $ \xibar_k $, as 
\beqa
\mubar_2 &=& 1 + \xibar_2 , \\
\mubar_3 &=& 1 + 3 \xibar_2 + \xibar_3 , \\
\mubar_4 &=& 1 + 6 \xibar_2 + 3 \xibar_2^2 + + 4\xibar_3 + \xibar_4 , \\
\mubar_5 &=& 1 + 10 \xibar_2 + 10 \xibar_3 + 15 \xibar_2^2 
+ 10 \xibar_2\xibar_3 + 5 \xibar_4 + \xibar_5 , 
\eeqa
also written as ``connected'' moments, 
\beq
\lexp N^{[k]} \rexp_c = \Nbar^k \, \xibar_k .
\eeq
The relations between the $ \mubar_k $ and the $ \xibar_k $ 
can be summarized in the generating functions 
$ M(t) $ and $ K(t) = \log M(t) $ \citep{F85}, 
\beqa
M(t) = \sum_{k=0}^\infty {1 \over k!} \, \Nbar^k \mubar_k t^k , \qquad
K(t) = \sum_{k=1}^\infty {1 \over k!} \, \Nbar^k \xibar_k t^k .
\eeqa
With the factors $1/k!$, $M$ and $K$ are sometimes called exponential 
generating functions of the moments $ \mubar_k $, $ \xibar_k $.
It is often found that the correlations vary with scale roughly 
as $ \xibar_k \propto \xibar^{k-1} $.
The hierarchical amplitudes $ S_k $ are defined by 
\beq
\xibar_k = S_k \, \xibar^{k-1} .
\eeq
The normalization of $ \xibar_k $ to $ S_k $ then removes 
much of the dependence of $ \xibar_k $ on scale.

Generating functions provide an interesting connection 
between discrete and continuous processes.
For a continuous variate $x$ with moment generating function 
$ M_c(t) = \lexp \e^{tx} \rexp $, the generator of a 
distribution of discrete counts $N$ for which $x$ 
is the local density is $ M_d(t) = M_c(\e^t-1) $ \citep{F85}.
This relation of generating functions provides directly 
the discreteness terms in equations~(\ref{disc2}--\ref{disc5}).
For discrete counts with probabilities $ P_N $, 
also useful is the probability generating function 
\beq
G(z) = \sum_{N=0}^\infty P_N \, z^N .
\eeq
For a discrete realization of an underlying continuous number density, 
$ G(z) $ is related to the exponential generating function of 
factorial moments by $ M(t) = G(t+1) $ \citep{F85,SS93}.

\section{Cell Counts on Large Scales: The Point Cluster Model}

Using the tools introduced in the previous section we can now 
construct the generating function of total number count in the 
point cluster limit.
On large scales, we expect that we can consider relatively compact 
clusters in their entirety to be either inside or outside of $V$.
The total number of galaxies in a volume is then the sum over 
all the clusters in the volume, 
\beq
N = \sum_{i=1}^{N_h} N_i , \label{sumNi}
\eeq
where the number of clusters $ N_h $ and the number of galaxies 
$ N_i $ in each cluster are chosen randomly and at first 
we take the cluster occupation numbers $N_i$ to be independent 
and identically distributed.
A similar sum over clusters arises in situations ranging from 
the distribution of particle multiplicities in hadron collisions 
at high energy accelerators \citep{F88,H94,T99} 
to the distribution of rainfall totals \citep{RICI87,C94,EF08}.

We can characterize the net count distribution directly for small 
counts and in general using the generating function $ G(z) $.
Let $ p_n $ be the probability of $V$ containing $n$ clusters,  
and let $ q_n $ be the probability that a cluster has $n$ members.
Because a cluster with no members is uninteresting, 
for simplicity we take $ q_0 = 0 $ here; 
for the case $ q_0 \neq 0 $, see Appendix A.
Then, to have no count, we must have no clusters; 
to have total count 1, we must have one cluster with one member; 
to have total count 2 we can have one cluster with 2 members 
or two clusters with 1 member, and so on.
The first several probabilities $ P_N $ that $V$ 
contains $N$ total galaxies are then 
\beqa
P_0 &=& p_0 \\
P_1 &=& p_1 q_1 \\
P_2 &=& p_1 q_2 + p_2 q_1^2 \\
P_3 &=& p_1 q_3 + 2 p_2 q_1 q_2 + p_3 q_1^3 \\
P_4 &=& p_1 q_4 + p_2 (2 q_1 q_3 + q_2^2) + 3 p_3 q_1^2 q_2 + p_4 q_1^3 \\
P_5 &=& p_1 q_5 + p_2 (2 q_1 q_4 + 2 q_2 q_3) \nonumber \\
&& + p_3 (3 q_1^2 q_3 + 3 q_1 q_2^2) + 4 p_4 q_1^3 q^2 + p_5 q_1^5 .
\eeqa
For these first few low order terms, $ P_N $ is the coefficient of $z^N$ 
in the composition of generating functions $ g_h \left[ g_i(z) \right] $.
This is the general result, as can be seen easily from the 
generating function for the total count probabilities $ P_N $, 
\beqa
G(z) &=& \Dlexp z^{N_1 + N_2 + \cdots + N_{N_h}} \Drexp 
= \Dlexp z{}^{N_i}\rexp{\!}^{N_h} \rexp \nonumber \\
&=& \lexp [g_i(z)]^{N_h} \rexp = g_h[ g_i(z) ] . \label{G(z)}
\eeqa
where the double angle brackets indicate an average over both 
the distributions of cluster occupancy and cluster number 
\citep[see][]{SS93}.

We can also compute moments directly and using generating functions.
The mean of $ N = \sum N_i $ is the number of haloes 
times the average occupation per halo,
\beq
\lexp N \rexp = \lexp N_h N_i \rexp = \Nbar_h \Nbar_i .
\eeq
The square $ N^2 = \sum N_i \sum N_j $ contains $ N_h $ terms 
with $ i = j $ and $ N_h (N_j-1) $ terms with $ i \neq j $, 
\beqa
\lexp N^2 \rexp &=& \lexp N_h N_i^2 + N_h(N_h-1) N_i N_j \rexp 
\nonumber \\
&=& \Nbar_h (\Nbar_i + \Nbar_i^2 \mubar_{2,i}) 
  + \Nbar_h^2 (1+\xibar_{2,h}) \Nbar_i^2 
\eeqa
Similar direct calculations give 
\beqa
\lexp N^3 \rexp &=& \lexp N_h N_i^3 + N_h (N_h-1) \, 3 N_i^2 N_j 
\nonumber \\
&&{}+ N_h (N_h-1) (N_h-2) N_i N_j N_k \rexp \\
\lexp N^4 \rexp &=& \lexp N_h N_i^4 
  + N_h (N_h-1) \, ( 4 N_i^3 N_j + 3 N_i^2 N_j^2 ) \nonumber \\
&&{}+ N_h (N_h-1) (N_h-2) \, 6 N_i^2 N_j N_k \nonumber \\
&&{}+ N_h (N_h-1)(N_h-2)(N_h-3)  \, N_i N_j N_k N_l \rexp \\
\lexp N^5 \rexp &=& \lexp N_h N_i^5 
  + N_h (N_h-1) \, ( 5 N_i^4 N_j + 10 N_i^3 N_j^2) 
  \nonumber \\
&&{}+ N_h^{[3]} \, (10 N_i^3 N_j N_k + 15 N_i^2 N_j^2 N_k) 
  \nonumber \\
&&{} + N_h^{[4]} \, 10 N_i^2 N_j N_k N_l 
+ N_h^{[5]} \, N_i N_j N_k N_l N_m \rexp , 
\eeqa
and from these, the discreteness corrected, 
connected moments of total count are 
\beqa 
\xibar_2 &=& \xibar_{2,h} + {\mubar_{2,i} \over \Nbar_h} \label{xi2} \\
\xibar_3 &=& \xibar_{3,h} + {3\mubar_{2,i} \xibar_{2,h} \over \Nbar_h} 
 + {\mubar_{3,i} \over \Nbar_h^2} \label{xi3} \\
\xibar_4 &=& \xibar_{4,h} + {6\mubar_{2,i} \xibar_{3,h} \over \Nbar_h} 
 + {(4 \mubar_{3,i} + 3 \mubar_{2,i}^2) \xibar_{2,h}  \over \Nbar_h^2} 
 + {\mubar_{4,i} \over \Nbar_h^3}  \label{xi4} \\
\xibar_5 &=& \xibar_{5,h} + {10 \mubar_{2,i} \xibar_{4,h} \over \Nbar_h} 
 + {(10 \mubar_{3,i} + 15 \mubar_{2,i}^2) \xibar_{3,h} \over \Nbar_h^2}
 \nonumber \\
&&{}\qquad + {(10 \mubar_{2,i} \mubar_{3,i} + 5\mubar_{4,i}) \xibar_{2,h} 
\over \Nbar_h^3} + {\mubar_{5,i} \over \Nbar_h^4} , \label{xi5}
\eeqa
etc.
Clearly, the effort and complexity increase at each order.
Identical results are obtained by the composition of 
generating functions in equation~(\ref{G(z)}).
The general term can be obtained from the generating function 
$K(t)$ for moments of total counts.
Using the relation $ M(t) = G(t+1) $, the composition of probability 
generating functions in equation~(\ref{G(z)} is also a composition 
of moment generating functions, 
\beqa
K(t) &=& \log[M(t)] = \log[G(t+1)] \nonumber \\
&=& \log\{g_h[g_i(t+1)]\} = \log\{g_h[M_i(t)]\} \nonumber \\
&=& \log\{M_h[M_i(t)-1]\} = K_h[M_i(t)-1] , \label{K(t)}
\eeqa
from which it is clear that $ \xibar_k $ continues to depend to all 
orders on the connected moments $ \xibar_{k,h} $ of halo number 
as the coefficients in $ K_h $ and the raw moments $ \mubar_{k,i} $ 
of the halo occupation distribution as the coefficients in $ M_i $.
The generating function in equation~(\ref{K(t)}) and the 
expressions for moments of counts in equations (\ref{xi2})--(\ref{xi5}) 
plus extension to higher orders 
constitute the main result of the point cluster model.
The point cluster results are independent 
of the internal details of halo profiles or concentrations.
The general expression for $ \xibar_k $ contains contributions from 
occupation number moments of order 1 through $k$ and halo correlations 
of order 1 through $k$; 
in $ \xibar_5 $, the first term arises from five objects in  five 
separate haloes, the last from occupancy five in a single halo, 
while other terms represent four haloes with occupancies (2,1,1,1); 
three haloes with occupancies (3,1,1) and (2,2,1); 
and two haloes with occupancies (3,2) and (4,1).
The numerical factors represent the number of equivalent halo assignments.
The sum of the combinatoric factors of a given $\xibar_{n,h} $ 
in the expression for $ \xibar_k $ are known as Stirling numbers 
of the second kind, $ S(n,k) $, the number of ways of putting $n$ 
distinguishable objects into $k$ cells with no cells empty \citep{SB91}.
Here they are produced from a generating function in a manner such 
that any term desired can be easily produced by an algebraic manipulator.

Some special cases are useful to consider.
For single-element clusters, $ N_i = 1 $ with probability 1, 
the occupation moments are $ \mubar_1 = 1 $ and $ \mubar_k = 0 $ 
for $ k \ge 2 $, galaxies are haloes and galaxy correlations 
are halo correlations, $ \xibar_k = \xibar_{k,h} $.
For Poisson occupation number, the occupation moments are all 
$ \mubar_{k,i} = 1 $, and the halo model expressions reproduce the 
discreteness terms of equations (\ref{disc2})--(\ref{disc5}).
This is the locally Poisson realization of a distribution 
with spatially varying $ n(\r) $.
For uncorrelated cluster positions, irreducible moments 
arise only from objects in the same halo; 
in this case the halo number $ N_h $ has a Poisson distribution, 
and the single-halo contribution to the count moment, 
\beq
\xibar_k^{1h} = {\mubar_{k,i} \over \Nbar_h^{k-1}} , \label{xiPois}
\eeq
is often called the Poisson term.
The composition of generating functions for a Poisson halo 
distribution was studied by \citet{S95a,S95b}.
The point cluster model varies from the halo clustering limit 
to the Poisson limit as a function of scale.
Typically, the two-point function behaves as 
$ \xibar(r) \sim r^{-\gamma} $ with $ \gamma \approx 2 $, 
and higher order correlations scale hierarchically, as 
$ \xibar_k = S_k \, \xibar^{k-1} $ with nearly constant $ S_k $. 
Since $ \Nbar $ grows as $ R^3 $, the dominant contribution 
to $ \xibar_k $ on large scales then comes from the halo correlation, 
$ \xibar_k \approx \xibar_{k,h} $, but 
on scales where $ \Nbar \xibar \la 1 $, 
the point cluster model gives the one-halo term 
in equation~({\ref{xiPois}).
In this regime total number count moments have hierarchical 
correlations, with $ S_k = \mubar_{k,i}/\mubar_2^{k-1} $.

Many common statistical models are constructed starting with 
a Poisson halo number distribution, so that  
$ \xibar_{k,h} = 0 $ for $ k \ge 2 $, and equation (\ref{xiPois}) 
holds exactly.
If the occupancy distribution is also Poisson, $ \mubar_{k,i} = 1 $, 
then $ S_k = 1 $ for all $k$ ($ S_1 = S_2 = 1 $ always), 
saturating constraints $ S_{2m} S_{2n} \geq S_{m+n}^2 $ 
arising from the Schwarz inequality; 
this is a realization of the minimal hierarchical model of \citet{F85}.
Other examples of compound Poisson distributions include the 
negative binomial distribution, which is the composition of a 
Poisson cluster distribution with a logarithmic occupation distribution 
\citep{S95b}, 
and the thermodynamic or quasi-equilibrium distribution of \citet{SH84}, 
which is the composition of a Poisson cluster distribution 
with a Borel occupation distribution \citep{S89,SS94,S95a}.

There is one generalization that is also useful, 
where the total number of objects is the sum of contributions 
from two independent populations, $ N = N_c + N_b $  
such as the sum of a strongly clustered population 
plus a weakly clustered ``background'' \citep[cf.][]{SP77}.
In this case the cumulant moments $ \xibar_k $ are simply additive, 
\beq
\Nbar^k \xibar_k \to \Nbar_c^k \xibar_{c,k} + \Nbar_b^k \xibar_{b,k} .
\eeq
If the background contributes to the total count, 
$ \Nbar = \Nbar_c + \Nbar_b $, but not to higher order moments, we have 
\beq
\xibar_k = {\Nbar_c^k \over (\Nbar_b + \Nbar_c)^k} \, \xibar_{c,k} 
= f_c^k \, \xibar_{c,k} , \label{f_c}
\eeq
where $ f_c $ is the fraction of clustered objects.
Although the correlations are diluted, this says that the 
amplitudes for $ k \ge 3 $ are amplified, $ S_k = S_{c,k} / f_c^{k-2} $.

\section{Numerical Results}

In this section we compare the model to statistics of galaxies and 
haloes identified within the setting of a single numerical simulation.
The sample we use is the same one analyzed in \citet[][hereafter CCT]{CCT07}, 
where many more details can be found. 
The simulation is performed with the adaptive mesh refinement (AMR) 
code {\RAMSES} \citep{Teyssier02}, assuming a standard $\Lambda$CDM 
cosmology with $\Omega_m=0.3$, $\Omega_\Lambda=0.7$, 
$ H_0 = 100 \, h \,{\rm km} \, {\rm s}^{-1} \, \Mpc^{-1} $ with 
$ h = 0.7 $, and a normalization $\sigma_8=0.93$, where $\sigma_8$ 
is the root mean square initial density fluctuations in a sphere of 
radius $ 8 \, h^{-1} \Mpc $ extrapolated linearly to the present time.
The simulation contains $512^3$ dark matter particles on the AMR grid, 
initially regular of size $512^3$, in a periodic cube of size 
$ L_{\rm box}=200 \, h^{-1} \Mpc $; 
the mass of a single particle is then $ 7.09 \times 10^9 \, \Msun $.
Additional refinement is allowed during runtime: cells containing 
more than $N_{\rm AMR}=40$ particles are divided using the 
standard AMR technique with a maximum of 7 levels of refinement. 

A halo catalog, $E_h$, and a ``galaxy'' (subhalo) catalog,
$E_h$, are extracted from the final state of the simulation 
using the publically available software {\adaptaHOP} \citep{APC04}; 
details of the procedure can again be found in CCT.
We use the number of dark matter substructures in each halo 
detected by {\adaptaHOP} as a proxy for the galaxy distribution.
A halo can contain one or more galaxies:  
a single component halo hosts one galaxy (or is its own substructure), 
and an $N$-component halo hosts $N$ galaxies.
The substructure distribution differs somewhat from that of 
galaxies \citep[see the discussions in CCT and][]{WCDK08}, 
but it provides a discrete number count distribution that is 
useful to test how the behavior of the discrete halo model 
differs from that of the continuous mass distribution.

Figure~\ref{fN} shows the distributions of halo mass $ f(\M) $
and of occupation number $ P_N $ for the full halo catalog.
The range of masses covers almost four decades; 
the largest halo contains 53 substructures.
Moments of these distributions give the occupation moments $ \mubar_k $ 
that appear in equations (\ref{xi2})--(\ref{xi5}) and the point 
halo hierarchical amplitudes $ S_k = \mubar_k / \mubar_2^{k-1} $.
For comparison, smooth lines in the figure show the Press--Schechter 
(solid line) and Sheth--Tormen (dotted line) mass functions, 
plotted for $ \delta_c = 1.50 $.
The Press--Schechter and Sheth--Tormen mass distributions provide a 
good representation of the mass function for $ M \ga 10^{12} \, \Msun $, 
rising with mass a little more weakly than $ 1/M $ towards small 
masses and with an exponential cutoff at large mass.
The number distribution behaves as a power law, $ P_N \sim 1/N^p $, 
with $p$ in the range 2.0--2.4.
The subclump finder {\adaptaHOP} identifies haloes as connected 
regions with density contrast larger than $ \delta > 80$ employing a 
standard SPH softening of the particle distribution with $N_{\rm SPH}=64$ 
neighbors \citep[see, e.g.,][]{Mo92}.
This, along with the mass resolution of the structures resolved by 
\RAMSES, controlled by the value of $N_{\rm AMR}$, leads to the rather 
soft small-$M$ cutoff on the halo mass function in Figure~\ref{fN} 
and also the low value of $ \delta_c $.

\begin{figure}
\includegraphics[width=\columnwidth]{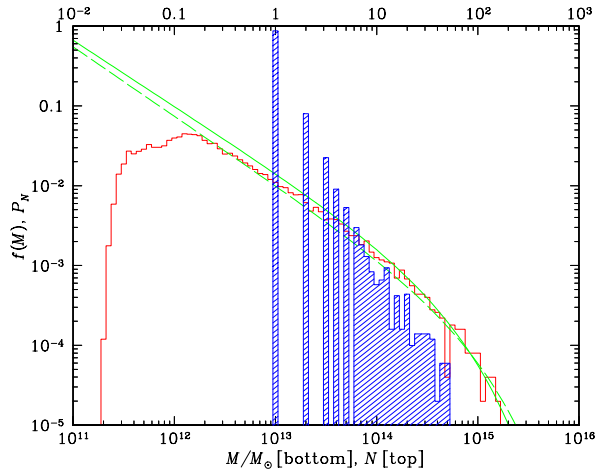}
\caption {Distribution of halo masses $ f(M) $ (histogram; bottom scale) 
as a function of $M$ and substructure occupation number probability $ P_N $ 
(shaded histogram; top scale) as a function of $N$.
The solid and dashed curves show the Press--Schechter 
and Sheth--Tormen mass functions.
\label{fN}}
\end{figure}

The full samples $E_h$ and $E_g$ contain 
50234 haloes and 64316 ``galaxies'', respectively.
Most haloes have a single component; the average number 
of substructures per halo is $\Nbar_i = 1.28$.
From these two parent catalogs, we apply various mass thresholdings
to extract subsamples from $E_h$ and $E_g$ that we 
denote $E_h (M_\min,M_\max)$ where $M_\min$ and $M_\max$, 
given in solar masses, correspond to minimum and maximum mass 
thresholds of the host haloes respectively.
We use these subcatalogs to test the variation 
of halo clustering with mass. 
The different realizations break down as follows: 
\begin{enumerate}
\item The full sample separated into ``light'' and ``massive'' 
  halo subsamples, $E_h(0,\infty) \equiv E_h \equiv 
  E_h(0,10^{14}) + E_h(10^{14},\infty)$, 
  and the substructure counterparts. 
\item A catalog of haloes with masses larger than $5 \times 10^{11} \Msun $, 
 which avoids the strongest rolloff at small mass, 
 separated likewise into two subsamples, $E_h(5 \times 10^{11},\infty) 
\equiv E_h(5 \times 10^{11},10^{14}) + E_h(10^{14},\infty)$,
  and the substructure counterparts;
\item A catalog of haloes with masses larger than $4 \times 10^{12} \Msun $, 
 which avoids essentially all of the rolloff at small mass, 
 separated likewise into two subsamples, $E_h(4 \times 10^{12},\infty) 
\equiv E_h(4 \times 10^{12},10^{14}) + E_h(10^{14},\infty)$,
  and the substructure counterparts.
\end{enumerate}
Table~\ref{table:catalogs} summarizes subcatalog information.

\begin{table}
\centering
\caption[]{
{Details of the catalogs extracted from \RAMSES.}
The two first columns give the halo mass range, 
$ M_\min < M_h < M_\max $ (in units of $\Msun$); 
the third and fourth columns give the number of objects 
in the halo and substructure catalogs; and 
the fifth and sixth columns give the average number of 
substructures and dark matter particles per halo.}
\begin{tabular}{ccrrrr}
\hline
$M_\min$ & $M_\max$ & $N_{\rm haloes}$ & $N_{\rm subs}$ & 
  ${\Nbar}_i^{\, \rm subs} $ & $ {\Nbar}_i^{\, \rm dm} $ \\
\hline
$0$                & $\infty$   &  50234 & 64316 &  1.28  &   967 \\
$0$                & $10^{14}$  &  49740 & 59203 &  1.19  &   636 \\
$5 \times 10^{11}$ & $\infty$   &  43482 & 57564 &  1.32  &  1109 \\
$5 \times 10^{11}$ & $10^{14}$  &  42988 & 52451 &  1.22  &   727 \\
$4 \times 10^{12}$ & $\infty$   &  11934 & 24918 &  2.09  &  3441 \\
$4 \times 10^{12}$ & $10^{14}$  &  11440 & 19805 &  1.73  &  2108 \\
$10^{14}$  & $\infty$   &    494 &  5113 & 10.35  & 34315 \\
\hline
\end{tabular}
\label{table:catalogs}
\end{table}

From the distribution of mass and the catalogs of haloes and subhaloes 
in the simulation we compute correlation statistics $ \xibar_k $ 
for $ k = 2 $--5.
Figure \ref{fig:xi2dm} shows the variance, or volume-averaged 
two-point correlation function $ \xibar_2 $, evaluated 
for spherical volumes of radius $R$ as a function of $R$, 
for dark matter, or mass (solid line), and for haloes (long-dashed line).
Dotted lines show the predictions of the point cluster model: 
the upper line for only the mass in haloes, and the lower line 
including mass not contained in haloes as an unclustered background, 
as in eq.~(\ref{f_c}), with $ f_c = \sum M_h / M_\tot = 0.36 $.
Finally, the short-dashed line includes two additional aspects 
from the full halo model, a modest relative bias factor $ b = 1.22 $ between 
mass and haloes on large scales, and the effects of resolved haloes, 
detailed in Section~5 below.

Panels in Figure \ref{fig:xi2abcd} show the second moment evaluated 
for substructure ``galaxies'' $ \xibar_{2,g} $ (solid lines) 
and for haloes, $ \xibar_{2,h} $ (long-dashed lines) 
for the four inclusive catalogs: haloes of all masses, haloes with 
$ M > 5 \times 10^{11} \, \Msun $, with $ M > 4 \times 10^{12} \, \Msun $, 
and with $ M > 10^{14} \, \Msun $, as identified in the caption.
Dotted curves show the predictions of the simple point cluster model.
Although it has shortcomings in detail, on scales of a few Mpc 
the point cluster model with no adjustments reproduces the trends 
with scale and from catalog to catalog to within a factor of two or so 
over four decades of correlation strength.
Dashed curves show a quantitative improvement with a very modest 
adjustment of parameters, relative bias factors of 1.2 or 1.3 
and occupation moments adjusted by a factor of 2, 
as given in the middle columns in Table~2.
For radius smaller than $ 1 \, h^{-1}{\rm Mpc} $ 
finite halo size starts to become important.
The mass threshold $ M > 5 \times 10^{11} \, \Msun $ 
removes only haloes containing a single substructure 
(the smallest halo containing two substructures has a mass 
$ 8 \times 10^{11} \, \Msun $), and so affects only the mean 
$ \Nbar = \lexp N \rexp $ but none of the higher factorial moments 
$ \lexp N^{[k]} \rexp $, just as for an unclustered background 
population as in equation~(\ref{f_c}).
Thus, in the regime where the normalized moment is large, 
$ \xibar_k \gg 1 $, it is simply rescaled, 
$ \xibar'_k / \xibar_k = (\Nbar/\Nbar')^k $.
This is apparent for the data plotted in Figure 2, where for the smallest 
cells $ \xibar_2 $ for the $ E_g(5\times 10^{11}) $ subsample 
is larger than that for the full $ E_g $ sample by a factor $ 1.249 $, 
very close to the number ratio $ (64316/57564)^2 = 1.246 $.
The next mass threshold, $ M > 4 \times 10^{12} \, \Msun $, 
removes doubly and also triply occupied haloes 
(the smallest halo containing three substructures has a mass 
$ 1.8 \times 10^{12} \, \Msun $), and so this threshold 
changes the shape of $ \xibar_2 $ and $ \xibar_3 $.

\begin{figure}
\includegraphics[width=\columnwidth]{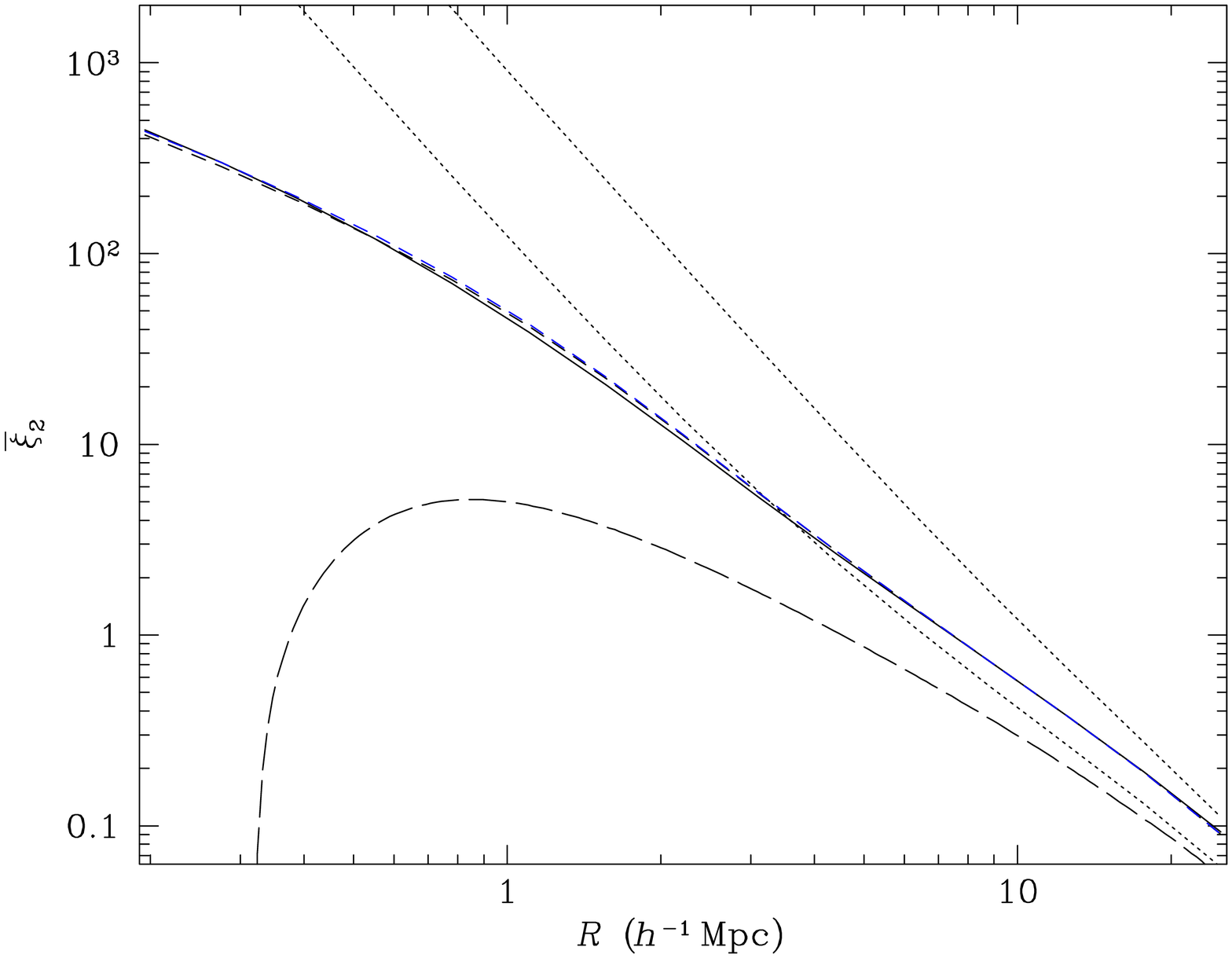}
\caption {Variance $ \xibar_2(r) $ for mass (solid line) and 
for halo number (long-dashed line) measured for spheres 
of radius $r$ in numerical simulations compared with the halo model.
Dotted lines show the point cluster model of equation  
(\protect{\ref{xi2}}); the upper line shows only the contribution 
of particles in haloes, while the lower dotted line includes 
particles not in haloes as an unclustered background.
The short-dashed line includes a bias $ b = 1.25 $ 
between mass and haloes on large scales 
and the effects of resolved haloes on small scales.
\label{fig:xi2dm}}
\end{figure}

\begin{figure}
\includegraphics[width=\columnwidth]{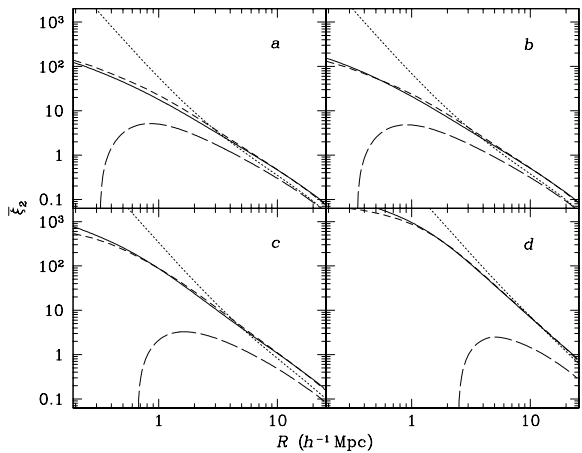} 
\caption {Cell count variance $ \xibar_2(r) $ in spheres 
of radius $r$ obtained from numerical simulations 
compared with the halo model.
Solid lines show $ \xibar_2 $ measured for galaxies (substructures) 
and long-dashed lines show $ \xibar_{_2,h} $ measured for haloes 
identified in the simulations for the four subcatalogs.
Results for all haloes are presented in panel ($a$), 
$ M > 5 \times 10^{11} $ in panel ($b$), 
$ 4 \times 10^{12} $ in panel ($c$), and $ 10^{14} \Msun $ in panel ($d$).
The dotted lines show the point cluster model of equation  
(\protect{\ref{xi2}}), and the short-dashed lines include 
bias and resolved haloes.
\label{fig:xi2abcd}}
\end{figure}

Panels in Figure~\ref{S3abcd} show the hierarchical amplitude 
$ S_3 $ for the four subhalo catalogs (solid lines),  
and for the corresponding halo catalogs (long-dashed lines).
Finite volume limitations are apparent at large scales, 
and the $ E_h(10^{14}) $ sample is not large enough 
for a reliable third moment on almost any scale.
Dotted lines show the naive point cluster model.
Again, the first mass cut $ M > 5\times 10^{11} \, \Msun $ 
removes only haloes containing a single substructure 
from the full catalog, changing only the mean count;  
and on the smallest scales the expected scaling 
$ S'_3/S_3 = \Nbar'/\Nbar = 57563/64316 = 0.895 $ 
is again satisfied.

\begin{figure}
\includegraphics[width=\columnwidth]{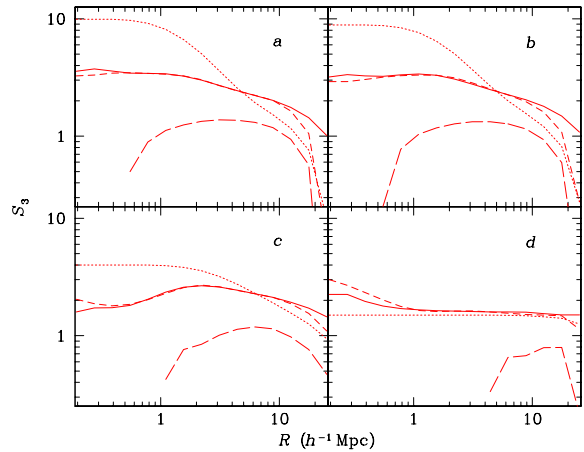} 
\caption {Hierarchical amplitudes $ S_3(r) $ from numerical simulations 
for substructure ``galaxy'' catalogs (solid lines) 
and for the corresponding haloes (long-dashed lines)
The dotted lines show the naive point cluster model, and 
the short-dashed lines include bias and resolved haloes.
Panels show the four different subcatalogs, 
as in Figure \ref{fig:xi2abcd}.
\label{S3abcd}}
\end{figure}

\begin{figure}
\includegraphics[width=\columnwidth]{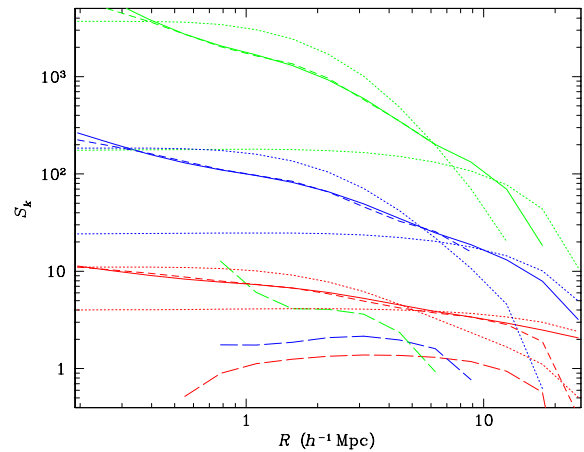} 
\caption {Hierarchical amplitudes $ S_k(r) $ for mass density 
from numerical simulations for $ k = 3 $, 4, and 5 (bottom to top).
Solid lines show $ S_k $ measured for mass; 
long-dashed lines show $ S_k $ for haloes.
Dotted lines show the naive point cluster model expressions in equations 
(\ref{xi3}--\ref{xi5}) and same results but adjusted 
for a weakly clustered background (not contained in haloes).
The short-dashed lines also include a bias factor on large 
scales and the effects of resolved haloes on small scales, 
as detailed in Section 5 below.
\label{SNdm}}
\end{figure}

\begin{figure}
\includegraphics[width=\columnwidth]{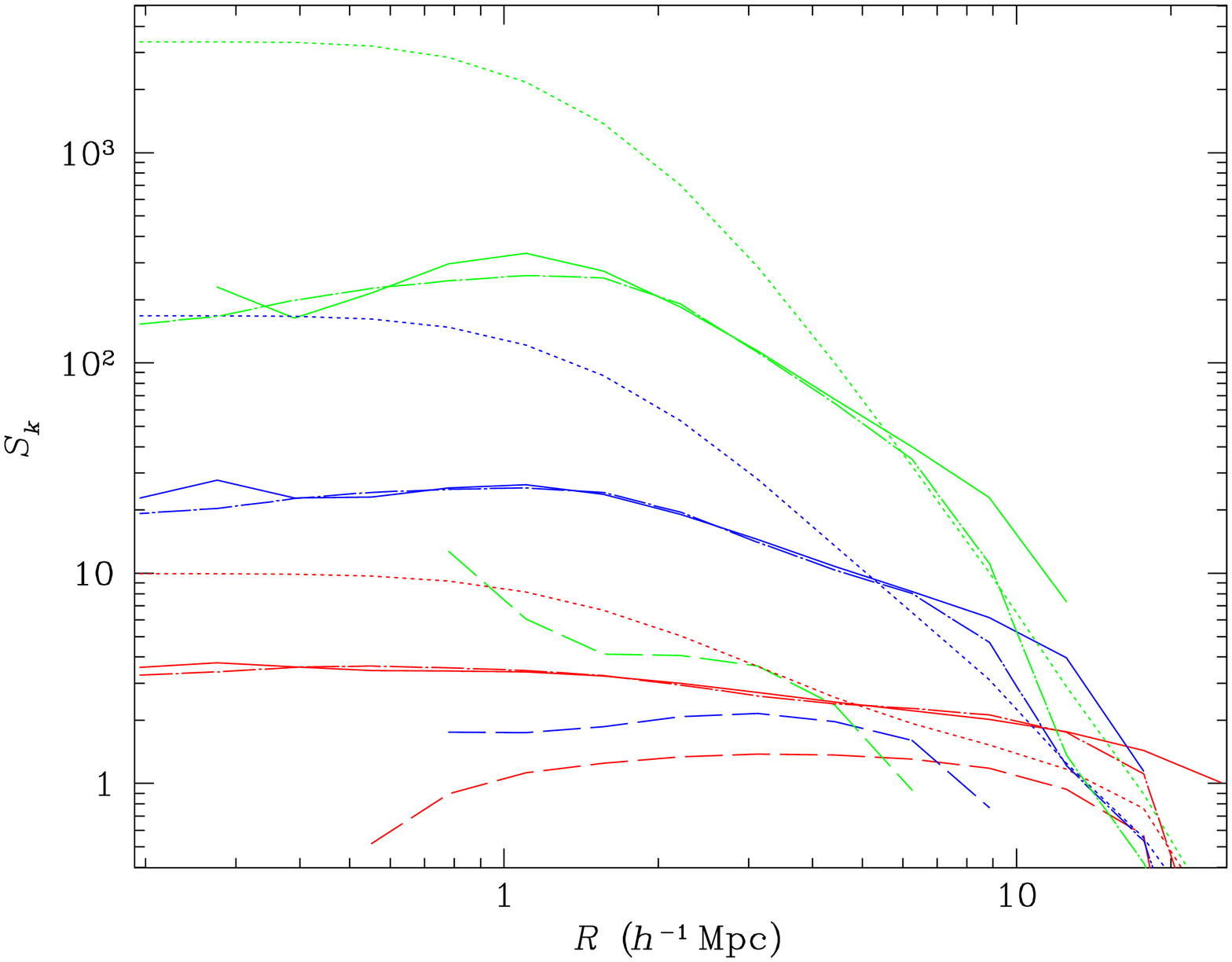} 
\caption {Hierarchical amplitudes $ S_k(r) $ for galaxy number count 
measured in from numerical simulations for $ k = 3 $, 4, and 5 
(bottom to top).
Solid lines show galaxy (substructure) $ S_{k} $, and 
long-dashed lines show halo $ S_{k,h} $.
Dotted lines show the point cluster model expressions in equations 
(\ref{xi3}--\ref{xi5}).
\label{Sn}}
\end{figure}

Figure \ref{SNdm} shows the amplitudes 
$ S_3 $, $ S_4 $, and $ S_5 $ for for dark matter (solid lines) 
and for haloes (long-dashed lines).
Figure \ref{Sn} shows the $ S_k $ for substructures (solid lines) 
and for haloes (long-dashed lines), for the entire $ E_h $ halo sample.
The naive point cluster model agrees with the simulations 
qualitatively but not quantitatively.
One possible explanation is that halo occupation is 
correlated with environment, and a modest adjustment 
of the point cluster parameters gives a good fit.
Table~2 shows the naive point cluster model result using 
occupation probabilities $ p_N $ and the halo mass function $ n(M) $
from the simulations, and also the result of adjusting fit parameters.
In the point cluster model, the parameters are factorial moments 
$ \mubar_k = \lexp N^{[k]} \rexp / \Nbar^k $ for galaxies 
and $ \mubar_k = \lexp M^k \rexp / \Mbar^k $ for mass, 
computed for the haloes identified in the simulation.
The quantity identified as ``$b$'' is the large-scale relative bias 
between galaxies and haloes, $ b^2 = \xibar_{2,g} / \xibar_{2,h} $.

\begin{table*}
\caption[]{Correlation parameters for ``galaxies'' and mass.
The first five columns show values in the simple point cluster model, 
in which all haloes are identical, 
using the simulation halo distributions $p_N $ and $ f(M) $; 
the next five columns show fit values (f); 
and the last five columns show values computed in the detailed 
halo model (m).}
\label{b3}
\begin{tabular}{@{}lrrrrrrrrrrrrrrrrr@{}}
\hline
Sample & $b/b_h$ & $\mubar_2$ & $S_3$ & $S_4$ & $S_5$ & 
$b/b_h^\fit$ & $\mubar_2^\fit$ & $S_3^\fit$ & $S_4^\fit$ & $S_5^\fit$ & 
$b/b_h^\model$ & $\mubar_2^\model$ & $S_3^\model$ & $S_4^\model$ & 
$S_5^\model $ \\
\hline
dark matter (mass) %
	& 1 & 4.00 & 7.18 & 93.0 & 1460 & 1.25 & 1.96 & 7.2 & 93 & 1460
	& 1.71 & 34.3 & 6.8 & 101 & 2450 \\
all haloes %
	& 1 & 1.35 & 9.92 & 168 & 3360 & 1.22 & 0.60 & 3.4  & 26 & 330 
	& 1.39 & 10.6 & 7.15 & 115 & 2980 \\
$ m > 5 \times 10^{11} \, \Msun $ %
	& 1 & 1.46 & 8.88 & 134 & 2410 & 1.23 & 0.70 & 3.4 & 25 & 300 
	& 1.37 & 6.95 & 5.24 & 59.5 & 1050 & \\
$ m > 4 \times 10^{12} \, \Msun $ %
	& 1 & 2.10 & 4.00 & 26.8 & 212 & 1.34 & 1.27 & 2.3 & 10 & 68 
	& 1.30 & 3.03 & 3.27 & 22.4 & 239 \\
$ m > 1 \times 10^{14} \, \Msun $ %
	& 1 & 1.50 & 1.50 & 2.93 & 6.59 & 1.36 & 1.09 & 1.7 & 4.0 & 13
	& 1.12 & 1.54 & 1.41 & 3.01 & 9.30 \\
\hline
\end{tabular}
\end{table*}

\section{The Full Halo Model}

To extend our understanding we turn to the context of the recently 
developed phenomenological halo model \citep{S00,MF00b,PS00,SSHJ01,CS02}, 
in which clustering on small scales is derived from the 
mass function, profiles, and clustering properties of dark matter haloes.
Numerical simulations have suggested haloes have a universal 
density profile \citep{NFW97,MQGSL99,NHPJFWSSQ04}, 
\beq
{\rho(r) \over \rhobar} = A \, u \bigl( {\,r\; \over \,r_s} \bigr) , 
\eeq
where the scale $ r_s $ and amplitude $A$ are functions of the halo mass.
In particular, $ r_s $ is related to the virial radius $ r_{200} $, 
within which the average density is 200 times the mean, by a 
``concentration parameter'' $c(\M)$, $ r_s = r_{200}/c $; 
this then also determines the amplitude $A$.
For a large cluster, say $ \M = 10^{15} \Msun $, the virial radius 
is about $ 3 \, h^{-1} {\rm Mpc} $; and with $ c \approx 6 $ 
the scale radius is $ r_s \approx 500 \, h^{-1} {\rm kpc} $.
Thus, at roughly Mpc and smaller scales we begin to resolve 
clusters, and we expect to have to replace the point cluster model 
with the full halo model.

The halo mass function is conveniently written as a function of 
the dimensionless overdensity $ \nu = \delta_c / \sigma(\M) $, 
where $ \delta_c $ is the threshold overdensity that leads 
to a collapsed halo, often $ \delta_c = 1.68 $, and $ \sigma^2(R) $ 
is the mean square mass fluctuation within a sphere of radius $R$ 
evaluated for the linearly evolved input power spectrum.
Specifically, 
\beq
\sigma^2(\M) = \int {\d^3k \over (2\pi)^3} \, P(k) \, W^2(kR) \,,
\label{sigma}
\eeq
where $ W(x)= 3(\sin x - x \cos x)/x^3 $ is the Fourier window
corresponding to a real-space top-hat window function, 
and $ m = 4\pi \rhobar R^3/3 $.
In terms of $\nu$, the density of haloes of mass $M$ is then written 
\beq 
{\d n\over \d\M} = {\d(\ln\,\nu) \over \d(\ln\M)} \, 
{\bar\rho \over \M^2} \, \nu f(\nu) .
\eeq
The Gaussian distribution function $ f(\nu) $ of \citet[][PS]{PS74}  
and the refinement of \citet[][ST]{ST99} have 
\beq
\nu f(\nu) = 2A \left[ 1 + (q\nu^2)^{-p} \right] 
\Bigl( {q\nu^2 \over 2\pi} \Bigr)^{1/2} \e^{-q\nu^2/2} .
\eeq
The normalization $A$ is chosen so that 
$ \int \d\nu \, f(\nu) = 1 $ and is independent of $q$.
The Press--Schechter function has $ q = 1 $, $ p = 0 $, $ A = \frac12 $; 
the Sheth--Tormen form has $ q = 0.707 $, $ p = 0.3 $, 
$ A \approx 0.32218 $. 

\subsection{One-Halo Term in the Point Cluster Limit}

In the point cluster limit of the full halo model, 
the sum over haloes in equation~(\ref{sumNi}) and the resulting 
composition of generating functions in equation~(\ref{G(z)}) 
remain true, but the calculation now includes an average over 
the distribution of halo masses as well as over halo occupation 
and halo count, both of which now differ with halo mass.
The resulting order-$k$ connected correlation function is again 
a sum of contributions from a single halo to $k$ different haloes, 
just as in equations~(\ref{xi2})--(\ref{xi5}), with the same coefficients.
For the one-halo term of the order-$k$ moment in the full halo model, 
the average over all haloes includes an average 
over the halo mass function $ dn/dm $, 
\beq
\xibar_k^{1h} = {\lexp N_i^{[k]} \rexp \over \lexp N_i \rexp^k } 
\to {\int \! \d m \, (\d n/ \d m) \, V \lexp N^{[k]}(m) \rexp \over 
\left[ \int \d m \, (\d n/ \d m) \, V \lexp N(m) \rexp \right]^k} , 
\eeq
where $ N(m) = N_i(m) $ is the occupancy of a halo of mass $m$.
The factor in brackets in the denominator is 
\beq
\int \! \d m \, {\d n \over \d m} \, V\lexp N(m) \rexp = 
\nbar_h V \lexp N_i \rexp = \Nbar_h \Nbar_i = \Nbar , \label{Nbar}
\eeq
where $ \nbar_h $ is the number density of all haloes, 
$ \Nbar_h = \nbar_h V $ is the mean number of haloes in $V$, 
and $ \Nbar_i $ is the average occupation 
over haloes of all masses; and the numerator is 
\beq
\int \! \d m \, {\d n \over \d m} \, V \lexp N^{[k]}(m) \rexp 
= \nbar_h V \lexp N_i^{[k]} \rexp = \Nbar_h \Nbar_i^k \mubar_{k,i} .
\eeq
The one-halo term of the full halo model thus produces the 
same result as the previous point cluster result, 
$ \xibar_k^{1h} = \mubar_{k,i} / \Nbar_h^{k-1} $, 
with occupation moment 
\beq
\mubar_{k,i} = {\int\!\d m \, (\d n/ \d m) \lexp N^{[k]}(m) \rexp 
 \, / \, \int\!\d m \, (\d n/ \d m) \, \over 
\, \left[ \int \d m \, (\d n/ \d m) \lexp N(m) \rexp \, / \, 
 \int\!\d m \, (\d n/ \d m) \right]^k } . \label{mubark1h} 
\eeq

We can compute occupation moments $ \mubar_k $ for mass from 
first principles by taking number $N$ to be proportional to mass 
(the number of hydrogen atoms or dark matter particles), $ N \propto m $ 
and $ N^{[k]} \propto m^k $ (with no discreteness terms).
From (\ref{mubark1h}), these are 
\beq
\mubar_k = {[\int\d m \,(\d n /\d m) \, m^k \,] \, 
 [\int\d m \,(\d n/\d m) \, ]^{k-1} 
\over [\int \d m \, (\d n / \d m) \, m \,]^k } , 
\eeq
with corresponding hierarchical amplitudes 
\beq
S_k = {\mubar_k \; \over \mubar_2^{k-1}} 
= { \bigl[ \int \d\nu \, f(\nu) \, m^{k-1} \bigr]
\bigl[ \int \d\nu \, f(\nu)  \bigr]^{k-2} \over 
\bigl[ \int \d\nu \, f(\nu) \, m \bigr]^{k-1} } \label{Skn}
\eeq
over the range of scales where the one-halo term dominates but 
haloes are not resolved.
In this case, results are determined entirely by the mass function, 
which in turn is related to the primordial power spectrum.
For a power-law power spectrum, with $ \nu = (\M/\M_1)^{(3+n)/6} $, 
and with the Press--Schechter and Sheth--Tormen forms of the 
mass functions, the integrals can be done analytically, giving 
\beq
S_k = { \; I(k) \;  [I(1)]^{k-2} \over [I(2)]^{k-1} } , 
\eeq
where 
\beqa 
I(k) &=& \Gamma \Bigl[  {3 (k-1)\over 3+n} + {1 \over 2} \Bigr] 
+ 2^{-p} \, \Gamma \Bigl[ {3 (k-1)\over 3+n} + {1 \over 2} - p \Bigr] , 
\eeqa
independent of $q$.
For a Poisson cluster distribution (on small scales cluster 
correlations are unimportant) and for the Press--Schechter mass function, 
this expression was also obtained by \citet{S96b}.
For the Press--Schechter mass function, which has $ p = 0 $, and for 
spectral index $ n = 0 $ this gives the particularly simple result 
$ S_k = (2k-3)!! $.
Results for power-law spectra are shown in Figure \ref{S_k(n)}, 
together with results from numerical simulations by \citet{CBH96} 
(plotted are the values of $ S_k $ measured at $ \xibar = 100 $, 
but values at $ \xibar = 10 $ or $ \xibar = 1 $ differ by less than 
the error bars).
The Sheth--Tormen mass function appears to agree poorly with 
the numerical results; 
this is one instance where the observed behavior seems to prefer 
the Press--Schechter form, at least for $n$ not too negative.
However, the Sheth--Tormen function is relatively more weighted 
towards smaller masses, and in numerical simulations there is 
always a smallest mass that can be considered.
Thus, we examine the results of a small mass cutoff in the integral, 
of $ 10^{-4} $ and $ 10^{-2} $ in units of the mass $ \M_1 $ 
at which $ \nu(\M_1) = 1 $.
A $ 10^{-4} $ cutoff mass has little effect on PS but is significant 
for ST, and a 0.01 cutoff has a significant effect on both.
In the simulations, the ratio of the particle mass to the 
non-linear mass is typically in the range 0.001--0.01, and 
the ST mass function with a moderate low-mass cutoff does agree 
with the simulations results, at least for $ -1 < n < 1 $.
As $n$ becomes more negative, all the halo model curves rise much 
more rapidly than the trend seen in the simulation results.
This may reflect an increasing difficulty in simulating 
negative values of $n$ \citep[cf.][]{JB98}.

\begin{figure}
\includegraphics[width=\columnwidth]{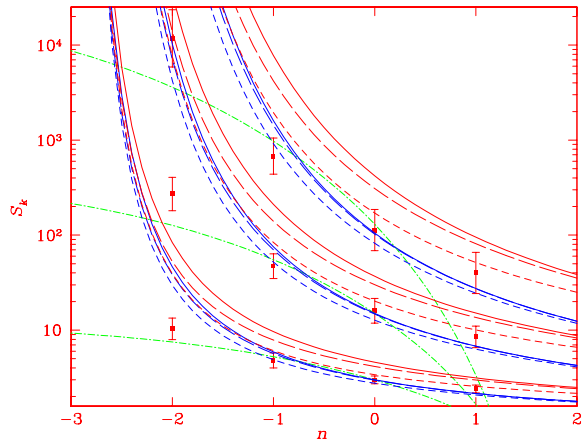} 
\caption {
Amplitudes $ S_k $ for $ k = 3 $, 4, 5 (bottom to top).
as a function of spectral index $n$ for power-law spectra.
For each $k$ curves are two sets of curves; 
the lower set (blue in color) shows results for the Press--Schechter 
mass function, and the upper set (red in color) for Sheth--Tormen, 
computed from eq.~(\ref{Skn}).
Solid lines show $S_k$ integrating over all masses; 
long-dashed curves have a lower mass cutoff $ \M > 10^{-4} $; 
and short-dashed curves have $ \M > 10^{-2} $, 
in units of the mass $ \M_1 $ at which $ \nu(\M_1) = 1 $.
The dot-dashed (green) curves show the predictions of 
hyperextended perturbation theory \citep{SF99}.
Symbols with error bars show results from numerical 
simulations \citep{CBH96}.
\label{S_k(n)}}
\end{figure}

For statistics of galaxy number counts, 
we must average over moments of halo occupation number, 
\beqa
S_k &=& {\bigl[ \int \d\nu  \, f(\nu) \lexp N^{[k]}(m) \rexp / m \bigr] 
\bigl[ \int \d\nu  \, f(\nu) \, \lexp N(m)\rexp /\M \bigr]^{k-2} \over 
\bigl[ \int \d\nu  \, f(\nu) \lexp N^{[2]}(m) \rexp / m \bigr]^{k-1} } 
\nonumber \\&&~
\eeqa
(the factor $ 1/\M $ remains from the PS or ST halo mass function).
In simulations, in general it is found the mean number of galaxies 
$ \lexp N(m) \rexp $ grows more slowly than linearly in mass.
Models have included a power-law, $ \lexp N(\M) \rexp  = (\M/\M_1)^\beta $, 
with $ \beta \la 1 $ and perhaps with a minimum mass cutoff $ \M_0 $; 
a broken two-power-law model \citep{BW02}; and a similarly behaved but 
smoothly interpolated function \citep{BWBBCDFJKL03}.
Substructures or subhalo occupation numbers exhibit a similar 
behavior, but perhaps with $ \beta \to 1 $ at high mass \citep{KBWKGAP04}.
Higher order correlations also require higher order moments of the halo 
occupation distribution, which are typically sub-Poisson at small $N$, 
with $ \lexp N(N-1)(m) \rexp < \lexp N (m)\rexp^2 $.
From semianalytic galaxy formation considerations, \citet{SSHJ01} 
extend to higher orders by assuming a binomial distribution, 
also used by \citet{KBWKGAP04}.
However, a representation of the galaxy number count distribution 
as a central galaxy plus a Poisson distribution of satellites 
\citep{BWBBCDFJKL03,KBWKGAP04,ZBWBBCDFKL05,ZW07,ZCZ07,ZZEWJ09} 
is becoming increasingly popular.
In this representation, both central and satellite distributions 
are characterized entirely by their means 
$ \lexp N_c \rexp = \Nbar_c $ and $ \lexp N_s \rexp = \Nbar_s $.
Since $ N_c $ takes on only the values 0 and 1, so that 
for any positive power $ \lexp N_c^p \rexp = \Nbar_c $, 
and since for a Poisson satellite distribution 
$ \lexp N_s^{[k]} \rexp = \Nbar_s $, 
the factorial moments of halo occupancy are 
\beq
\lexp N^{[k]} \rexp 
= \Nbar_s^k + k \, \Nbar_c \Nbar_s^{k-1} .
\eeq
The central object can be modeled as a sharp 
or smoothed step function \citep{BWBBCDFJKL03}, and 
\citet{ZBWBBCDFKL05} present expressions for moments with 
parameters extracted from simulations.
A main import of all models is that moments of occupation number 
grow more slowly than linearly with mass, 
a behavior that we model as the simpler form 
$ N(\M) \sim \M^\beta $ with $ \beta \la 1 $.
With the Sheth--Tormen mass function and with no small-mass cutoff, 
$ S_k $ for number counts is again a ratio of $\Gamma$-functions, 
\beq
S_k = { I_\beta(k) \,  \left[I_\beta(1)\right]^{k-2} 
\over \left[I_\beta(2)\right]^{k-1} } , 
\eeq
where now 
\beq
I_\beta = \Gamma \Bigl[ {3\beta (k-1) \over (3+n)} + {1 \over 2} \Bigr] 
+ 2^{-p} \, \Gamma \Bigl[ {3\beta (k-1) \over (3+n)} + {1 \over 2} - p \Bigr] .
\eeq

We can use the Poisson model to obtain the occupancy 
probability distribution averaged over all haloes.
For a Poisson distribution with mean $ \mu(m) $, 
the probability $ p_N $ for a halo of mass $m$ to contain 
$N$ galaxies (or $N$ satellite galaxies) is 
$p_N = \mu^N \, e^{-\mu} / N! $.
Averaged over the power-law portion of the Press--Schechter 
mass function $ dn/dm \sim \nu/m^2 $, with the integral cut off by 
the Poisson exponential $ e^{-\mu} $ 
before the exponential cutoff in the mass function is reached, 
the probability $ p_N $ of $N$ objects in any halo scales as 
\beq
p_N \propto {\left[ N + {(3+n) / 6\beta} - 
1 - {1 / \beta} \right]! \over N!} , 
\eeq
where $ \nu \sim m^{(3+n)/6} $ and $ \mu(m) \sim m^\beta $.
As $N$ becomes large, this behaves as a power law, 
\beq
p_N \sim N^{-r} , \qquad 
r = 1 + {1 \over \beta} - {(3+n) \over 6\beta} . 
\eeq
For $ n \approx -2 $ and $ \beta \la 1 $ the exponent is near $ r = -2 $, 
a good approximation to the distribution plotted in Figure~1.
For Sheth--Tormen the power is shifted by $ 2p(3+n)/6\beta $, 
or by about 0.1.

\subsection{Resolved Haloes}

For small volumes we can no longer take haloes as point objects, but 
must take into account the distribution of objects within a halo.
In the full halo model, the one-halo contribution to 
the $k$-point function $ \xi_k^{1h} $ for mass is a convolution 
of halo profiles \citep{MF00b}, 
\beqa
\xi_k^{1h} = {\int \d m \, (\d n/\d m) \, m^k 
\int \d^3 r' u(y'_1) \cdots u(y'_k) \over \left[ \int \d m \, 
(\d n/\d m) \, m \int \d^3r' \, u(y') \right]^k} , 
\label{xik1h} 
\eeqa
where the position $\r'$ of the halo centre runs over 
all space, $ y'_i = |\r_i-\r'|/r_s $, and the scaled halo profile 
$ u(r) $ is normalized to unit integral.
From equation~(\ref{xik1h}), the volume-averaged correlation is then 
\beqa
\xibar_k^{1h} = 
{\int \d m \, (\d n/ \d m) \,m^k \int \d^3 r' \, [F(r')]^k 
\over \left[ \int \d m \, (\d n/ \d m) \, 
m \int \d^3r' \, F(r') \right]^k} , \label{xi1h} 
\eeqa
where $F(r')$ is the portion of the total volume of a halo 
centred at $\r'$ that lies within $V$, 
\beq
F(r') = \int_0^R \d^3r \; u \Bigl( {\r-\r' \over r_s} \Bigr) . 
\label{F}
\eeq
Note that the integrand is a function of $r/r_s $, 
and since the scale radius $r_s$ depends on mass, 
the from factor $F$ is in general also a function of halo mass.
From equations~(\ref{xi1h}) and (\ref{F}) we can recover the point 
cluster model: if a volume is much larger than a halo size, 
$ R \gg r_s $ for all haloes, then $ F(r') $ is very small unless the halo 
itself is within $V$, in which case the integral then contains 
the entire halo contents.
In this limit and with unit normalization, $ F \to 1 $ for 
$\r'$ in $V$ and $ F \to 0 $ for $ \r' $ outside $V$.
Then, the integral over $\r'$ is just a factor of $V$, 
and we recover the point cluster model.

For resolved haloes in moments of discrete galaxy counts 
we consider first the second count moment $ \xibar_2 $.
Let a halo contain $N$ objects, and let $N'$ be the number of these 
objects that are contained within $V$.
Then $N' = \sum N_i$, where either $ N_i = 1 $ with probability $p_i$ 
if object $i$ is counted or $ N_i = 0 $ if object $i$ is not, 
and the second moment is 
\beqa
\lexp N' (N'-1) \rexp &=& 
\Lexp \tsum_{i=1}^N N_i \, \bigl( \tsum_{j=1}^N N_j -1 \bigr) \Rexp 
\nonumber \\
&=& \Lexp \tsum_{i \neq j} N_i N_j 
 + \tsum_{i=1}^N N_i^2 - \tsum_{i=1}^N N_i \Rexp . 
\eeqa
But since $N_i$ takes on the values 0 or 1, 
$ N_i^2 = N_i $ and the last two terms cancel, leaving 
the sum only over distinct objects 
\beq
\lexp N'(N'-1) \rexp = \Lexp \tsum_{i \neq j} N_i N_j \Rexp 
= \lexp N(N-1) \rexp p^2 .
\eeq
If object positions within a halo are uncorrelated, the probability 
$p$ that an object within a given halo is located within the volume $V$ 
is just the fraction $F$ of the halo that is within $V$, 
form factor in equation~(\ref{F}), the same for all objects and 
independent of the halo occupation $N$, 
\beq
\lexp N' (N'-1) \rexp = \lexp N(N-1) \rexp \lexp F^2 \rexp .
\eeq
This agrees with the usual practice, to distribute the average 
pair count $ \lexp N(N-1) \rexp $, weighted by the square 
of the halo profile form factor $ \lexp F^2 \rexp $, 
\beq
\mubar_2 = {\int \d m \, (\d n/ \d m) \, \lexp N(N-1) \rexp \, 
\lexp F^2 \rexp / \, \nbar_h \over \left[ \, \int \d m \, (\d n/ \d m) 
\, \lexp N(m) \rexp / \, \nbar_h \, \right]^2 } , 
\label{mubar_2}
\eeq
where the volume-averaged form factor is 
\beq
\lexp F^k \rexp = {1 \over V} \int_0^\infty d^3r' \left[ F(\r') \right]^k .
\label{F^k}
\eeq
In the position space formulation symmetry over all particles is manifestly 
maintained in the form-factor integrals, without need to introduce 
the approximation $ W_{12} \approx W_1 W_2 $.
The form factor $F$ does not appear in $ \Nbar $ in the denominator 
of equation (\ref{mubar_2}), since, as can be easily seen by changing 
the order of integration, $ \lexp F \rexp = 1 $.
The calculation for a halo occupation distribution consisting 
of a central object plus $ N_s = N-1 $ satellites yields
\beq
\lexp N'^{[2]} \rexp = \lexp N_s^{[2]} \rexp \lexp F^2 \rexp 
+ 2 \lexp N_s \rexp \lexp F \, F_c \rexp , 
\eeq
where $ F_c = 1 $ for $ r < R $ and vanishes otherwise.
Extending to general $k$, we obtain  
\beq
\lexp N'^{[k]} \rexp = \lexp N^{[k]} \rexp \, \lexp F^k \rexp \label{Fk}
\eeq
with no central object, or 
\beq
\lexp N'^{[k]} \rexp = \lexp N_s^{[k]} \rexp \lexp F^k \rexp 
+ k \lexp N_s^{[k-1]} \rexp \lexp F^{k-1} \, F_c \rexp , \label{Fkc} 
\eeq
with a central object.
The last term could contain a factor $ \lexp N_c \rexp $ 
if this is not 1.
The form-factor-corrected halo occupation moment is then 
\beq
\mubar_{k} = {\int \d m \, (\d n/ \d m) \, \lexp N^{[k]}(m) \rexp \, 
\lexp F^k \rexp / \, \nbar_h \over \left[ \, \int \d m \, (\d n/ \d m) 
\, \lexp N(m) \rexp \, / \, \nbar_h \, \right]^k , } 
\label{mubar_k}
\eeq
modified as in equation~(\ref{Fkc}) for a central object.

For moments of dark matter mass, a reasonably good representation 
of the numerical results is obtained using the NFW profile, 
but for substructures this is not the case.
The substructure profile was seen in \citet{DMS04} to follow 
roughly an isothermal profile, and we have studied using 
the isothermal sphere profile also.
The measurements of \citet{DMS04} and our own do not 
provide enough statistics to infer a mass dependence of the 
concentration parameter, and so we use a constant value 
$ c = 10 $ that gives reasonable results on small scales.
Figure \ref{FF} shows the volume-averaged form factor for 
$ k = 2 $--5 for NFW haloes (solid lines) and for the isothermal 
sphere profile (long-dashed lines), both with $ c = 10 $.
Curves are plotted as a function of $ Y = R/r_s $, where 
$ r_s = r_{200} / c $.
As expected, the form factor goes to 1 at large scale 
and falls rapidly for small $R$, 
where only a small fraction of a halo is sampled.
Note that $ \lexp F^k \rexp \le \lexp F^n \rexp $ if $ k < n $.
The integral converges to 1 on large scales, the point cluster 
regime, but falls rapidly for $ Y < c $.
In equation~(\ref{mubar_k}), for fixed $R$, 
this factor decreases rapidly for increasing mass.

Figure \ref{sn1h} shows the form-factor corrected, one-halo 
$ S_k = \mubar_k / \mubar_2^{k-1} $, 
normalized by its value in the point-cluster limit, 
as a function of $R$, for $ k = 3 $, 4, and 5 
(bottom to top; different orders $k$ offset for clarity).
On small scales, smaller than a few Mpc, this shows the effect 
of resolved haloes.
The result depends on both halo profile and on the 
distribution function: 
solid lines show NFW haloes averaged over mass; 
long-dashed lines show the same haloes averaged over number; 
short-dashed lines show the isothermal profile with $ c = 10 $; 
and dotted lines show isothermal profiles with concentration 
$ c(m) $ as for NFW.
On large scales halo profile shape has no effect, 
but on small scales the differences for different 
profiles and weightings are substantial.

From the halo model can extract small-scale behaviors 
of the correlations $ \xibar_k $, which can be different for galaxies 
and for mass.
The concentration parameter plays a critical role in the result.
For scale invariant spectra \citep[c.f.][]{DP77} we expect 
$ c(\M) \sim M^{-\alpha} $, with $ \alpha = (3+n)/6 $ 
\citep[in][ this parameter is $\beta$]{MF00a}.
For $\Lambda$CDM, over our relatively small range in mass we 
also take the concentration parameter to scale as a power of mass, 
$ \alpha \approx 0.11 $ or $ 0.12 $ \citep{BKSSKKPD01,ZJMB03}, 
corresponding to an effective $ n \approx -2.3 $.
As above, let the number of objects in a halo grow with mass as 
$ \lexp N^{[k]} \rexp \sim m^{k\beta} $ 
(for statistics of mass $ \beta = 1 $).
Finally, let $ \d n/\d\M \sim \nu^{p'} /\M^2 $ as $ \M \to 0 $, 
where $ \nu \sim \M^{(3+n)/6} $ and 
$ p' = 1 $ for PS and $ p' = 1-2p = 0.4 $ for ST 
\citep[in][ this parameter is $\alpha$]{MF00a}.
Then, ignoring the exponential factor in $ \d n/\d\M $ on small scales 
where $\nu$ is small and changing integration variable from $\M$ to 
$ Y = R / r_s = c R/r_{200} $ in equation~(\ref{mubar_k}), 
we see that $ \mubar_k $ scales as 
\beq
\mubar_k \sim \left[ R^{1/(\alpha + 1/3)} \right]^%
{ [k \beta + p' (3+n)/6 - 1]} 
\eeq
and the $k$-point function $ \xibar_k = \mubar_k / \Nbar_h^{k-1} $ 
scales as $ R^{-\gamma_k} $, with 
\beqa
\gamma_k &=& 
{3 \over 1+3\alpha} \left[ 3 (k-1) \alpha + k (1-\beta) \right] 
- {(3+n) p' \over 2 (1 + 3 \alpha)} \\
&=& (k-1) {3(5 - 2\beta + n) \over 5+n} + {6(1-\beta) \over 5+n} 
- {(3+n)p' \over 5+n} , 
\eeqa
independent of the shape of the halo profile.
For $ \beta = 1 $ this is the same as the result obtained 
by \citet{MF00a} (beware a change of notation) for mass, 
and for $ \beta = 1 - \epsilon $ is the result obtained by 
\citet{SSHJ01} for galaxy number.
This is of the hierarchical form only for $ p' = 0 $, 
which is not true for either of the PS or ST mass functions, 
and for $ \beta = 1 $.
Departures from hierarchical scaling in the small-$R$ 
behavior of $ S_k $ grow with $k$, 
\beq
S_k \sim R^{(k-2) \, \Delta \gamma} , 
\eeq
where 
\beq
\Delta \gamma = { (3+n)p' - 6 (1-\beta) \over 2 (1+3\alpha)} 
= {(3+n)p' - 6 (1-\beta) \over 5+n} \label{Dg}
\eeq
($ \Delta \gamma \approx -0.26 $ for the choices 
$p' = 0.4 $, $ \beta \approx 0.8 $, $ n \approx -2 $).
The presence of ever higher powers of $ \xibar_2 $ in 
$ S_k = \xibar_k/\xibar_2^{k-1} $ emphasizes any scaling defects 
in $ \xibar_2 $.
An interesting alternative normalization is 
\beq
S'_k = {\xibar_k \over \xibar_{k-1}^{(k-1)/(k-2)}}
= {S_k \over S_{k-1}^{(k-1)/(k-2)} . }
\eeq
Departures from scaling in $ S'_k $ decrease with $k$ for 
$ k \ge 3 $, as 
\beq
S'_k \sim R^{\Delta \gamma /(k-2) } 
\eeq
for the same $ \Delta \gamma $ given in equation~(\ref{Dg}).
 
\begin{figure}
\includegraphics[width=\columnwidth]{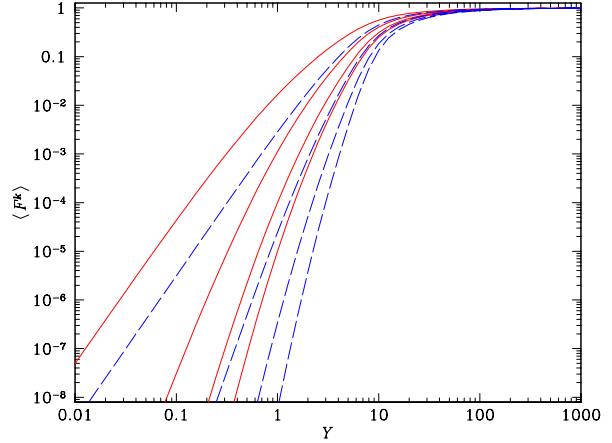}
\caption {Volume-averaged form factor $ \lexp F^k \rexp $ for 
$ k = 2 $--5 (upper left to lower right), 
as a function  of $Y = R/r_s $.
Solid lines shows result for NFW profile; 
long-dashed lines show isothermal profile, both with $ c = 10 $.
\label{FF}}
\end{figure}

\begin{figure}
\includegraphics[width=\columnwidth]{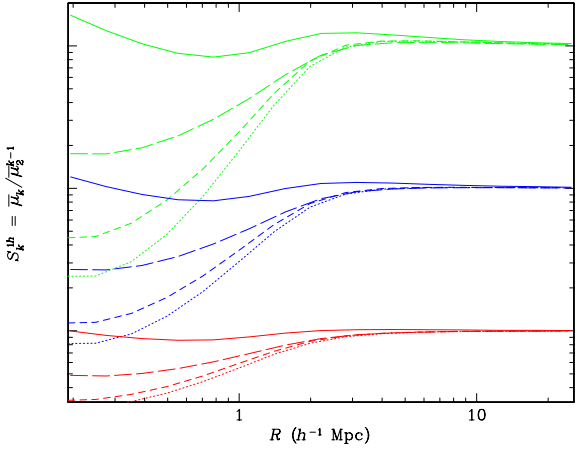}
\caption {Form-factor corrected one-halo $ S_k $ as a function of $R$, 
normalized to its point cluster value, 
for $ k = 3 $, 4, and 5 (bottom to top; different $k$ offset for clarity).
Solid lines show $ S_k $ for NFW profiles weighted by mass $ \M^k $; 
long-dashed lines show NFW profile weighted by number $ N^{[k]} $; 
short-dashed lines show isothermal profile with $ c = 10 $ weighted 
by number; and dotted lines show isothermal profile with $ c = c(m) $.
\label{sn1h}}
\end{figure}

\subsection{Multiple-Halo Terms}

Terms that involve objects in multiple haloes also depend on 
correlations among haloes.
In the perturbative regime, 
halo correlations have bias factors that are functions 
of the halo masses, and higher order correlation functions 
also involve higher order bias parameters \citep{FG93}; 
for instance, the halo three-point function is 
\beqa
\xi_{3,h} &=& b(m_1) b(m_2) b(m_3) \, \xi_{3,\rho}(r_{12},r_{13},r_{23}) 
 \nonumber\\
&&\qquad {} + b_2 \, [\xi_{2,\rho}(r_{12}) \xi_{2,\rho}(r_{13}) 
+ \hbox{cyc. (123)}] ,
\eeqa
where the $ \xi_\rho $ are correlation statistics for 
the underlying (primordial) density distribution.
As a function of mass, the linear bias factor found for PS by 
\citet{MJW96,MJW97} and adapted for the ST halo mass function 
\citep{ST99,CMB03} is 
\beq
b = 1 + {q \nu^2 - 1 \over \delta_c} 
 + {2p \over \delta_c} {1 \over [1 + (q \nu^2)^p]} , 
\eeq
with further refinements for small mass suggested by \citet{J99}.
Higher order functions also require higher order bias parameters 
\citep{MJW97,SSHJ01}, 
\beqa
b_2 &=&{8 \over 21\delta_c} \left\{ q \nu^2 - 1 
 + {2p \over [1 + (q \nu^2)^p]} \right\} \nonumber \\
&& +  {1 \over \delta_c^2} \left\{ q^2 \nu^4 - 3 q \nu^2 
 + {2 p (q \nu^2 + 2p - 1)  \over [1 + (q \nu^2)^p]}  \right\} ,
\eeqa
etc.
Higher order bias terms $b_3$, etc., vanish when integrated over 
the full halo mass function.
Even with a low mass cutoff or with different mass or number 
weightings we expect that they remain  generally small; 
and so we will drop them from now on \citep[but see][]{ABL08}.

We exhibit in detail the $k$-halo and two-halo contributions to 
$ \xibar_k $ in the full halo model.
The $k$-halo contribution to $ \xibar_k $ is 
\beqa
\Nbar^k \xi_k^{kh} &=& \prod_{i=1}^k 
 \int \d\M_i \, {\d n \, \over \d\M_i} \lexp N (\M_i) \rexp \\
&& \times \int \d^3 r'_1 \, u(y'_1) \cdots \d^3 r'_k \, u(y'_k) 
\, \xi_{k,h}(\r'_1 , \dots , \r'_k) , \nonumber 
\eeqa
where $ \Nbar $ is as given in equation~(\ref{Nbar}).
Ignoring non-linear bias terms, so that in terms of the underlying 
density correlation $ \xi_{k,\rho} $ the halo correlation function is 
$ \xi_{k,h} = b(\M_1) \cdots b(\M_k) \, \xi_{k,\rho} $, 
the volume-averaged correlation becomes 
\beqa
\Nbar^k \xibar_k^{kh} &=& \prod_{i=1}^k
\int \d\M_i \, {\d n\, \over \d\M_i} \, 
	b(\M_i) \lexp N(\M_i) \rexp \\
&& \times \int \d^3 r'_1 \, F(r'_1) \cdots \d^3 r'_k \, F(r'_k) 
 \, \xi_{k,\rho}(\r'_1 , \dots , \r'_k) .  \nonumber 
\eeqa
In the point cluster limit on large scales, for which $ F = 1 $ 
for $ \r' $ in $V$ and $ F = 0 $ for $ \r' $ outside $V$, 
this gives 
\beq
\xibar_{k,h}^{kh} = \bbar^k \, \xibar_{k,\rho} 
= {\bbar^k \over \bbar_h^k} \, \xibar_{k,h} , 
\eeq
with an occupation-number weighted bias factor, 
\beq
\bbar = {\int \d\M \, (\d n/\d \M) \, \lexp N(\M) \rexp \, b(\M) 
\over \int \d\M \, (\d n/\d \M) \lexp N(\M) \rexp } , \label{bbar}
\eeq
The halo correlation function $ \xibar_{k,h} $ is 
$ \xibar_{k,h} = \bbar_h^{\,k} \, \xibar_{k,\rho} $, 
with a bias factor weighted only by the halo mass distribution, 
\beq
\bbar_h = {\int \d\M \, (\d n/\d \M) \, b(\M) 
\over \int \d\M \, (\d n/\d \M) } . \label{bbar_h}
\eeq
Factors of mass or number weight greater contributions at higher masses, 
where $b(m)$ takes on larger values, so in general $ \bbar > \bbar_h $; 
galaxies are more strongly correlated than haloes on large scales, 
though only by a small amount.
Ratios of integrals $ \bbar/\bbar_h $ over the Sheth--Tormen mass 
function with $ \lexp N \rexp \propto \M $ for mass and 
$ \lexp N \rexp \propto \M^\beta $ with $ \beta = 0.8 $ 
for number are listed in Table~2.

Similarly, we can write intermediate terms.
The two-halo contribution to $ \xibar_k $ 
is a sum of terms of the form 
\beqa
\Nbar^k \xibar_k^{2h} &=& {s \, k! \over k_1! \, k_2!}
\int \d\M_1 \, {\d n \, \over \d\M_1} \; \d\M_2 \, {\d n \, \over \d \M_2} 
\lexp N_1^{[k_1]} \rexp \lexp N_2^{[k_2]} \rexp \, 
\nonumber \\
& \times & \!\int \, \d^3 r'_1 \, \d^3 r'_2 \, 
[F(r'_1)]^{k_1} [F(r'_2)]^{k_2} \, b(\M_1) b(\M_2) \, \xi_2(r'_{12}) , 
\nonumber \\
\qquad 
\eeqa
where $ k = k_1 + k_2 $.
(If $ k_1 = k_2 $ there is an additional symmetry factor of 
$ s = \frac12 $ because the partition and its complement are identical; 
the generating function gives all combinatoric factors automatically.)
On large scales, where the halo size is insignificant and the form 
factors take the value $F=1$ over essentially the entire volume $V$, 
the full two-halo term is thus the sum over partitions 
\beq
\xibar_{k}^{2h} = {s \, k! \, \over k_1! \; k_2!} \, 
{\bbar_{k_1} \over \bbar_h} {\bbar_{k_2} \over \bbar_h} 
\, {\mubar_{k_1} \, \mubar_{k_2} \, \xibar_{2,h} \over \Nbar_h^{k-2}} ,  
\label{xi2h}
\eeq
where $ \bbar_k $ is weighted by $ \lexp N^{[k]} \rexp $, 
\beqa
\bbar_k &=& {\int \d\M \, (\d n/\d \M) \, 
\lexp N^{[k]}(\M) \rexp \lexp F^k \rexp b(\M) \over 
\int \d\M \, (\d n/\d \M) \, \lexp N^{[k]} \rexp \lexp F^k \rexp} . 
\label{bbar_k}
\eeqa
For moments of mass, the factors $ \lexp N^{[k]}(\M) \rexp $ 
become $ \M^{k} $.
Weighted by different factors of number or mass, 
the bias parameters $ \bbar_k $ will in general be different 
from $ \bbar = \bbar_1 $ defined in equation~(\ref{bbar}); 
the lower mass limit for the integral also increases 
for higher order moments.
On small scales, where haloes are resolved, the halo size, 
and thus the factor $ F(r') $, depend also on halo mass: 
the mass and position integrals cannot be factored or simplified.
However, since $ F \leq 1 $, the expression in equation~(\ref{xi2h}) 
is an upper limit to the two-halo contribution, 
and even without the form factors the two-halo contribution 
is dominated by the one-halo term given in 
equation~(\ref{xi1h}) as $ R \to 0 $.
Extension to other intermediate orders follows similar lines.

\section{Discussion}

We have studied the behavior of cell count moments, including 
the variance $ \xibar_2 $ and the hierarchal amplitudes $ S_k $ 
for $ k = 3 $, 4, and 5, in the context of the halo model, 
and we have compared the model with results of numerical 
simulations for statistics mass and of galaxy (substructure) 
number counts identified in the same simulation.
Expressions (\ref{xi2})--(\ref{xi5}) constitute the halo model 
predictions for the two-, three-, four-, and five-point functions; 
a composition of generating functions for the halo number 
and halo occupancy distributions, as presented in equation~(\ref{K(t)}), 
produces automatically the halo model result at general order, 
including all terms and combinatoric factors.
The naive, point-cluster form of the model with identical haloes 
is easily generalised to include averages over a distribution of 
halo masses and over positions within resolved haloes.
The general form of the naive point cluster model results continues 
to hold, with the addition of a modest bias, of a factor of two or 
less, on large scales, and form factors that reflect shapes 
of resolved haloes on small scales.
With these components, the halo model is able to reproduce in 
quantitative detail statistical moments for mass and for substructure 
samples whose densities vary by a factor of one hundred.

On scales greater than of order a few Mpc, theoretical predictions 
are well represented in the point cluster version of the halo model.
The point cluster model results range from a biased realization 
of halo correlations on large scales to intermediate scales, 
for which $ \Nbar \xibar \la 1 $, where single halo contributions 
dominate, but haloes are still unresolved.
In this limit the results of the halo model are independent of details 
such as halo profile, asphericity, and concentration parameter.
Intermediate results are robust; the variance 
$ \xibar_2 = \mubar_2 / \Nbar_h^2 $ steepens, approaching $ r^{-3} $, 
an effect seen in scaling studies \citep{HKLM91,PD96}; 
and the amplitudes $ S_k $ are constant, 
as in the plateau seen in scale free models by \citet{CBH96}.
These results are independent of profile shape or bias.
The halo model with Press--Schechter or Sheth--Tormen mass function 
allows us to compute from first principles values for the 
hierarchical amplitudes $ S_k $ for scale-invariant models with 
power spectrum $ P \sim k^n $.
As shown in Figure~\ref{S_k(n)}, the halo model predictions 
are sensitive to the mass function and to mass cutoffs.
For scale-free models with initial spectrum $ P(k) \sim k^n $, 
the halo model reproduces the general trends of $ S_k(n) $.
Disagreements for more negative $n$ are probably an indication 
of the difficulty of simulating these spectra.

The largest halo has $ r_s \approx 500 \, h^{-1} {\rm kpc} $; on scales 
smaller than this, we must include the effects of finite halo size.
Resolved haloes introduce scale-dependent form factors in the 
$ \mubar_{k,i}(R) $, as in Section~4.
Analysis of resolved haloes is made substantially more efficient by 
analytic expressions for the form factors, contained in Appendix~C.
Small-scale results suggest that the profile shape is different 
for mass and for substructures.
In our limited efforts we have not found a profile shape that allows 
us to fit the shape of $ S_k $ on all scales in the resolved halo regime.
For that matter, we do not really know that a universal profile shape 
applies for the distribution of galaxies within haloes of different size; 
A possible explanation is that tidal disruption leads to 
no universal profile that applies on all scales; 
or, the halo model picture itself may be oversimplistic.
Nevertheless, on large scales our simulation and model results 
seem to have the potential to agree with observations \citep{RBM06}.

As observations become more and more precise, so it is increasingly  
important to be able to model clustering statistics with precision.
This appears to be possible for mass on both large and small scales.
On large scales, perturbation theory (biased linear theory) 
is accurate to better than 1\%.
On small scales, where statistics are dominated by tightly bound, 
high-density collapsed haloes, 
using published forms for the mass function $ f(m) $ and the 
concentration parameter $ c(m) $ with no attempt to optimize, 
the halo model reproduces the variance for our simulation 
again to within a few percent.
This is suggestive but not in itself a proof of the halo model; 
history has shown that there may be many constructs that lead to 
the same two-point function; thus, it is a nontrivial result 
that the halo model also reproduces with accuracy the higher order 
correlation functions on small and large scales as well.
There are somewhat larger deviations on intermediate scales, 
where the halo model predictions are too large by 5--10\%, 
in a direction that is only made worse by including higher order 
perturbative corrections.
This is a regime where the halo model seems to be least likely 
to be valid, where there is a significant amount of inhomogeneously 
clustered mass not contained in spherical haloes; 
another interesting possibility in this regime is 
renormalized perturbation theory \citep{CS06}.
The halo model predictions also match very well number count statistics 
on both large and intermediate scales, the point cluster regime, 
across all the different subcatalogs with different mass thresholds.
This is perhaps not a surprise, since there is no ``background'' 
population of objects outside haloes; 
objects in haloes account for all the objects there are.
However, such precision does not seem to be within reach on small scales.
The results we present use an isothermal profile with fixed $ c = 10 $, 
but this is at best only a first approximation.
With a small number of haloes, we do not know the profile shape, 
although it seems that the NFW profile does not work, 
and we do not know how the halo radius or concentration parameter 
depends on mass.
This may be the result of using substructures instead of 
galaxies in a full hydrodynamic simulation; 
substructures in high density regions, 
may be tidally disrupted \citep{WCDK08}.

Halo model statistics computed over mass and number distributions 
taken from the simulation work well.
It is in principle possible to compute correlations from first principles, 
starting with a primordial power spectrum, using the Sheth-Tormen 
halo mass function and a prescription such as a Poisson satellite number.
Application to scale-free simulations with initial spectrum $ P \sim k^n $ 
gives plausible results for $ S_k(n) $, at least for $n$ not too negative, 
once taking into account finite simulation resolution.
In practice, the predicted relative bias factors  $\bbar_g/\bbar_h$ 
do not quite match the numerical results.
but this is probably due to finite volume effects.
In particular, the halo five-point function is barely detected.

In the end, on small scales there are substantial differences between the 
discrete statistics of number counts and the continuous statistics of of mass.
The distribution function of halo occupation number has a behavior 
different from that of the distribution of halo mass, 
and factorial moments of discrete counts behave differently 
than moments of mass, even if the mean occupation 
number itself were a faithful tracer of total mass, 
all which contribute to differences in $S_k$, 
both in value and in shape as a function of scale, to the extent that 
it is not clear that the concept of bias between galaxy and mass 
statistics, even a non-linear bias, is a useful concept.

It may sometimes seem that with a halo profile shape, mass function, 
concentration parameter, and asphericity all to be specified, 
the halo model is infinitely adjustable.
However, on intermediate and large scales much of this freedom 
disappears, and the model depends only on the compounding of 
statistics.
In the halo model calculation, we see that the overall size of the 
correlation function $ \xibar_k $ or the amplitude $ S_k $ 
is determined by moments $ \mubar_k = \lexp \M^k \rexp $ of mass 
or factorial moments $ \mubar_k = \lexp N^{[k]} \rexp $ 
of halo occupation number; 
while details of shape on small scales provide information 
on the halo profile, $ \lexp F^k \rexp $.
That the model can reproduce in detail the measured $ S_k $ 
for $ k = 3 $--5, simultaneously for both mass and number, 
and can handle probabilities as well as moments, 
is a nontrivial success.

\section*{Acknowledgments}

We thank David Weinberg for many helpful comments and suggestions.
I.S. is grateful to NASA for support from grant NNG06GE71G.
J.N.F. thanks the IAP for hospitality during this work.
Parts of this work were clarified in discussions at the 
Aspen Center for Physics workshop on the Halo Model.
This research has made use of NASA's Astrophysics Data System.



\appendix

\section{Empty Haloes}

We show in this Appendix that for every composite distribution $ P_N $ 
that includes empty haloes with a nonzero probability $ q_0 \neq 0 $, 
there is another with $ q'_0 = 0 $ that produces the same $ P_N $.
Thus, excluding (or including) empty haloes does not impose a physical 
restriction on the resulting occupation distribution $ P_N $.

Suppose we start with a distribution with $ q_0 \neq 0 $.
Then, the total count probabilities will include contributions 
from many clusters with no occupancy, 
\beqa
P_0 &=& p_0 + p_1 q_0 + p_2 q_k^2 + p_3 q_0^3 +\cdots \\
P_1 &=& p_1 q_1 + 2 p_2 q_1 q_0 + 3 p_3 q_1 q_0^2 + \cdots \\
P_2 &=& p_1 q_2 + p_2 (q_1^2 + 2 q_2 q_0) 
 + p_3 (3q_2 q_0^2 + 3 q_1^2 q_0) + \cdots \\
P_3 &=& p_1 q_3 + p_2 (2 q_1 q_2 + 2 q_3 q_0) \nonumber \\
 &&\qquad  + p_3 (q_1^3 + 6 q_1 q_2 q_0 + 3 q_3 q_0^2) + \cdots 
\eeqa

We can easily create an occupancy distribution with no empty haloes 
while maintaining the same relative probabilities by defining a new 
set of probabilities $ q'_n $ such $ q'_0 = 0 $ and 
$ q'_n = q_n / (1-q_0) $ for $ n > 0 $.
This distribution has the generating function
\beq
g'_i(z) = {g_i(z) - q_0 \over 1 - q_0} .
\eeq
Note that since $ g_i(0) = q_0 $, this gives $ g'_i(0) = 0 $, 
and $ g'_i(1) = g_i(1) = 1 $.

We can then modify the halo occupation number probability $ p'_n $ 
in what turns out to be a sensible way to produce in the end the 
same $ P_N $.
With $ q'_0 = 0 $ we must have $ P_0 = p'_0 $, so we take 
\beq
p'_0 = \sum_{n=0}^\infty p_n q_0^n = g_h(q_0) .
\eeq
Next, to have $ P_1 = p'_1 q'_1 $. 
\beq
p'_1 q'_1 = {p'_1 q_1 \over 1-q_0} = \sum_{n=0}^\infty n p_n q_1 q_0^{n-1} 
= q_1 \left. {\d g_h(z) \over \d z} \right|_{z = q_0}, 
\eeq
we take $ p'_1 = (1-q_0) (\d g_h/ \d z)|_{q_0} $.
Similarly, to have 
\beqa
P_2 &=& p'_1 q'_2 + p'_2 q'^2_1 \nonumber \\
&=& q_2 \sum n p_n q_0^{n-1} + q_1^2 \frac12 \sum n(n-1) p_n q_0^{n-1} , 
\eeqa
we take $ p'_2 = \frac12 (1-q_0)^2 (\d^2g_h/\d z^2)|_{q_0} $.
These $p'_n $ follow from the generating function 
\beq
g_h'(z) = g_h[ (1-q_0) z + q_0 ] .
\eeq
The coefficient of each term in the expansion is a 
sum of products of positive numbers with positive coefficients, 
and so $ p'_n \ge 0 $; and $ g_h'(1) = g_h(1-q_0 + q_0) = 1 $, 
so that each term must satisfy $ p'_n \le 1 $ and 
the distribution is properly normalized.
The composition of these two modified distributions then gives  
\beq
G'(z) = g_h'[g_i'(z)] = g_h[ (1-q_0) g_i'(z) + q_0] 
= g_h[ g_i(z) ]  , 
\eeq
and so the same $ P_N $, as desired.

For the case of a Poisson cluster number distribution, 
\citet{F88} shows that the revised $ p'_n $ again are a Poisson 
distribution with mean $ \Nbar' = \Nbar (1-q_0) $.
The general case has essentially the same interpretation.
The continuum (discreteness corrected) moments, 
generated by $ M(t) = G(1+t) $, follow from  
\beqa
M'_h(t) &=& G'(1+t) = G[(1-q_0)(1+t) + q_0)] \nonumber \\
&=& G[1 + (1-q_0) t] = M[(1-q_0) t ] .
\eeqa
Thus, continuum moments are scaled by a factor $ (1-q_0)^n $ 
which absorbed in the mean $ \Nbar'_i = \Nbar_i (1-q_0) $, 
leave the correlations $ \mubar_n $ unchanged: 
the $q'_k$ are a discrete realization of the same 
underlying number density field.
In a sense, this is the equivalent of including an unclustered 
background, as in equation~(\ref{f_c}).

In general $ q_0 $ can be mapped to any value 
$ 0 < \alpha < 1 $ by the transformations 
\beq
g'_i(z) = {g_i(z) - \alpha \over 1 - \alpha} ,  \qquad 
g'_h(z) = g_h[ (1-\alpha) z + \alpha ] , 
\eeq
and it remains true that the generating function of total 
count probabilities is unchanged, 
$ G(z) = g'_h[g'_i(z)] = g_h[g_i(z)] $.

\begin{table*}
\caption[]{
{Parameters used to perform the count-in-cells measurements.}
The first line gives the inverse scale in units of the simulation box 
size $L_{\rm box}/\ell$; the smallest and the largest scales, 
$ L_{\rm box}/\ell =1024 $ and $ L_{\rm box}/\ell = 8 $, correspond
to $\ell=0.2 \, h^{-1} \Mpc $ and $\ell = \, 25 h^{-1} \Mpc $, respectively.
The second line gives the size of the grid of sampling cells, 
$N_{\grid}$, used to perform the measurements at a given $\ell$ 
for the full dark matter sample, \RAMSES; 
$N_{\grid}=2048$ means that $2048^3$ cells were used, 
corresponding to a minimum possible value of $P_N$ 
of the order of $1.16\times10^{-10}$.
The third line identifies the count-in-cell measurement method used 
for each scale under consideration, T for oct-tree walk, 
F for FFT, and D direct assignment.
The fourth and fifth lines give $N_{\grid}$
and the method for all the other samples.}
\begin{tabular}{lccccccccccccccc}
\hline
$L_{\rm box}/\ell$  &   1024  & $512\sqrt{2}$ & 512  & $256\sqrt{2}$ &
256 & $128\sqrt{2}$ & 128 & $64\sqrt{2}$ & 64 & $32\sqrt{2}$ &
32 & $16\sqrt{2}$ & 16 & $8\sqrt{2}$ & 8  \\
\hline

\RAMSES  &&&&&&&&&&&&&&& \\

$N_{\grid}$  &  2048  & 2048  &  2048  & 2048  & 2048  &
2048  & 1024  & 1024  & 1024  & 1024 & 512  & 512  & 512  & 512  & 512 \\

method   &   T    &   T  &   T   &   T   &   T   &  T   &   T  & %
T   &   T   &   T  &   T  &  T  &  F  &   F  &   F \\

\hline
Others &&&&&&&&&&&&&&& \\
$N_{\grid}$  &  1024 &  1024  & 1024  & 1024 &  1024  &  1024 %
& 1024  & 1024  & 1024  & 1024 & 512 &  512 &  512 &  512 &  512 \\

method  &  D  &  D  &  D  &  D  &  D  &  D  &  D  & %
D  &  D  &  D  & D  &  D  &  D  &  D & D \\
\hline
\end{tabular}
\label{table:measurements}
\end{table*}

\section{Algorithms for Computing the Count-in-Cell Distribution Function}

In this Appendix we detail how the count-in-cell distribution function
$ P_N(\ell) $ is estimated in these samples. 
There exist many efficient ways to measure this function in cubical
cells \citep[e.g.,][]{Szapudi98,SQSL99,BSCBBDGPS06}. 
The problem is however more intricate for spherical cells 
of radius $\ell$, which we prefer to use in this paper, 
since the analytical calculations 
are much easier to derive for these latter.
Although it is rather usual and fair to approximate spherical cells 
with cubical cells of same volume with a small form factor correction 
\citep[e.g.,][and references therein]{Szapudi98}, 
we prefer here to avoid this approximation.
Then, the two most common ways of measuring function $P_N(\ell)$ 
for spherical cells are
\begin{enumerate}
\item {\em The FFT method:} it consists in assigning the particles
to a grid of size $N_{\grid}$ using e.g. nearest grid point or
cloud-in-cell interpolation \citep[e.g.,][]{HE81}, Fast
Fourier Transform (FFT)
the corresponding density distribution, multiply the result by the Fourier
transform of the top hat filter in Fourier space, and then Fourier
transform back to estimate function $P_N(\ell)$. Obviously the FFT
method is valid only if the cell size is much larger than the size
of a mesh element. 
\item {\em The direct assignment method:} it consists in creating
a list of candidate cells positioned on a regular pattern of size
$N_{\grid}$, then on scanning the list of particles and assigning
them to each cell when relevant to augment the corresponding count.
This method does not suffer the defects of the FFT, so can be used
even for very small scales, but can become prohibitive at large scales,
when the network of cells starts to significantly self-overlap 
(i.e. a particle is assigned to a large number of cells). 
\end{enumerate}
Naturally, a good choice consists in using the direct assignment
method at small scales and the FFT method at large scales. 
However, the very large number of particles, $512^3$, in our full 
{\RAMSES} dark matter called for an additional algorithmic improvement, 
valid at intermediate scales.
Our implementation, {\tt CountKD}, is a code based on a 
decomposition of space using standard oct-tree technique, 
similarly as in \citet{SQSL99}, but for spherical cells.
The oct-tree decomposition allows one to approximate a
spherical cell with a coverage of cubes of varying size, this latter
decreasing (by factors of 2) when approaching the cell boundary. 
At some point, when
maximum allowed level of refinement is reached (or when there is only
one particle per oct-tree cell), the position of particles themselves
is used to decide if they belong to the cell in consideration or not.
Obviously, this method is efficient only if there are sufficiently
many particles per cell, otherwise it is faster to perform direct
assignment as explained above.
For this work, we did not bother to find the optimal compromise between
direct assignment, oct-tree walk and FFT method and performed the
measurements as described in Table~\ref{table:measurements}. 
The main goal was simply to reach a reasonable level of accuracy 
to sample correctly the large $N$ tails of function $P_N(\ell)$ in a 
reasonable amount of CPU time while avoiding as much as possible 
the FFT method which is sensitive to the pixelization of the data.

\section{Form Factor Integrals}

In this Appendix we present analytic expressions for the form factors 
$ F(x;y,c) $, the fraction of a halo with concentration parameter 
$c$ and centre at $ r/r_s = x $ that is contained within a 
spherical volume of radius $ R/r_s = y $.
First, note that the volume ${\cal V}$ of the intersection of two spheres 
of radii $a$ and $b$ whose centres are separated by distance $d$ is 
\beq
{\cal V}(a,b,d) = \cases {
I(a,b,d) , & $ |a-b| \le d \le a+b $, \cr
\noalign{\medskip}
0 , & $ d > a+b $, \cr
\noalign{\medskip}
\displaystyle{\frac{4\pi}{3}} \, \min(a^3,b^3) , & $ d \le |a-b| $, \cr } 
\eeq
where the shared volume $ I(a,b,d) $ of spheres that partially overlap is 
\beqa
I(a,b,d) &\equiv& \frac{\pi}{d} \left[ \frac{1}{12} d^4 
	- \frac{1}{2} d^2 (a +b^2) \right. \nonumber \\
&&{}\qquad\qquad \left. + \frac{2}{3} d (a^3 + b^3) 
	- \frac14 (a^2 - b^2)^2 \right] .
\eeqa
Next, note that a decreasing profile $\rho(r)$ can be modelled as 
a sum of step functions, truncated at an outer radius $r_\max$.
This means that the profile is composed of an ensemble of spheres of 
decreasing densities $\rho_i$ and increasing radii $r_i$,
$i=0,\dots,N$, with $r_0 = 0$, $r_N = r_\max$.
Then, for $\rho(r) = \rho_{i}$ for $ r_{i-1} < r \le r_i $, 
we can write 
\beq
\rho(r)=\rho_N - \sum_{r_i > r} \Delta\rho_i, \quad
\Delta\rho_i\equiv \rho_{i} - \rho_{i-1}. \label{rho}
\eeq
Now, let us compute the mass contained within radius $R$ 
of the origin for a halo centred at a distance $d$, 
\beq
M(d,R) = \int_{r < R} \d^3r \, \rho({\bm d}+{\r}).
\eeq
From equation~(\ref{rho}), 
the calculation reduces to the sum of intersections 
of spheres with appropriate weights, 
\beq
M(d,R) = \rho(r_\max) \, {\cal V}(R, r_\max,d) 
 - \sum_i {\cal V}(r_i,R,d) \Delta \rho_i .
\eeq
In the continuous limit, we obtain the final expression
\begin{eqnarray}
M(d,R) &=& \rho(r_\max) {\cal V}(R,r_\max,d) \nonumber \\
&&{} - \int_{|R-r| < d < r+R}^{r < r_\max} 
 I(r,R,d) \, \frac{\d\rho}{\d r} \, \d r \nonumber \\
&&{} - \int_{d < |r-R|}^{r < r_\max} \frac{4\pi}{3} \, 
  \min(R^3,r^3) \, \frac{\d\rho}{\d r} \, \d r .
\end{eqnarray}
We apply this in particular to the NFW and isothermal sphere profiles.

\subsection{Convolution of the NFW profile}
The truncated NFW profile writes, in scaled units 
\beq
\rho_{\NFW}(r) = \cases{ 
\displaystyle {1 \over r(1+r)^2}, & $ r \le c $, \cr
\noalign{\smallskip}
0 , & $ r > c  $ .\cr } \label{eq:NFWnaive} 
\eeq
The calculation of the integral gives
\beqa
M_\NFW(x) &=& -\left\{ I(r,y,x)\rho(r) 
  - {2\pi (x+1) \over x} \ln(1+r) \right. \nonumber \\
 && \left. \quad + {\pi r \over x} - {\pi r \over x} \left[
  \frac{y^2-(x+1)^2}{1+r} \right]\right\}_{r=r_-}^{r=r_+} \nonumber \\
 && - \left\{ \frac{4}{3}\pi r^3 \rho(r) - 
  4\pi \left[ \frac{1}{r+1} + \ln(r+1) \right] \right\}_{r=0}^{r=r_1}
  \nonumber \\
 && - \left\{  \frac{4}{3}\pi y^3 \rho(r) \right\}_{r=r_-}^{r=c}
 + \rho(c) {\cal V}(c,y,x) ,  \label{eq:rhoconv}
\eeqa
where $ \rho = \rho_\NFW $, 
$ x = d/r_s $, $ Y = R/r_s $, $ c = r_\max/r_s $, and 
\beqa
r_+ &=& \min(c,x+y), \label{r-} \\
r_- &=& \min(c,|x-y|) , \label{r+} \\
r_1 &=& \min[c,\max(y-x),0)] .\label{r1}
\eeqa
Evaluated for $ y > c $ and $ x = 0 $ this gives the total mass, 
\beq
M = 4\pi {[(1+c) \ln(1+c) - c ]\over (1+c)} , 
\eeq
and the normalized NFW form factor is then 
\beq
F_\NFW(x,Y,c) = {M_\NFW(x;Y,c) \over 4\pi [(1+c) \ln(1+c) - c]/(1+c) } . 
\eeq

\subsection{Convolution of the Isothermal profile}
The isothermal profile writes, in scaled units
\beq
\rho_{\ISO}(r)= \cases { 
\displaystyle {1 \over 1+r^2}, & $ r \le c $, \cr
\noalign{\smallskip}
0 , & $ r > c $ . \cr } 
\eeq
The calculation of the integral gives
\begin{eqnarray}
M_{\ISO}(x) &=& 
- \Bigl\{ I(r,x,d) \rho(r) - 2\pi (r-\arctan r) \nonumber \\
&& \quad + {\pi \over 2x} \left[ (x^2 - y^2 - 1) 
\log(1 + r^2) + r^2 \right] \Bigr \}_{r=r_-}^{r=r_+} \nonumber \\
&& -\left\{ \frac{4\pi}{3} r^3 \rho(r) 
  - 4\pi (r - \arctan r) \right\}_{r=0}^{r=r_1} \nonumber \\
&&-\left\{ \frac{4}{3}\pi y^3 \rho(r) \right\}_{r=r_-}^{r=c} 
+ \rho(c) {\cal V}(c,y,x) , 
\end{eqnarray}
where $ \rho = \rho_\ISO $ and $ r_+ $, $ r_- $, and $ r_1 $ 
are as in equations~(\ref{r-})--(\ref{r1}).
Evaluated for $ y > c $ and $ x = 0 $ this gives the total mass, 
\beq
M = 4\pi (c - \arctan c) ,  
\eeq
and the normalized form factor for the isothermal profile is 
\beq
F_\ISO(x,Y,c) = {M_\ISO(x;Y,c) \over 4\pi (c - \arctan c) } . 
\eeq

\end{document}